\documentclass[english]{article}
\usepackage[T1]{fontenc}
\usepackage[latin9]{inputenc}
\usepackage{geometry}
\usepackage{array}
\usepackage{longtable}
\usepackage{units}
\usepackage{mathtools}
\usepackage{multirow}
\usepackage{amsmath}
\usepackage{amssymb}
\usepackage{graphicx}
\usepackage{rotating}
\usepackage{natbib}

\makeatletter

\providecommand{\tabularnewline}{\\}

\newcommand{\lyxaddress}[1]{
	\par {\raggedright #1
	\vspace{1.4em}
	\noindent\par}
}

\makeatother

\usepackage{babel}
\begin{document}
\title{Estimation of Shannon differential entropy: An extensive comparative review}
\author{Mbanefo S. Madukaife$^{1,2}$ and Ho Dang Phuc$^{2}$}
\maketitle

\lyxaddress{$^{1}$Department of Statistics, University of Nigeria, Nsukka, Nigeria; $^{2}$Department
of Probability and Mathematical Statistics, Institute of Mathematics,
Vietnam Academy of Science and Technology, Hanoi, Vietnam}
Corresponding Author email: mbanefo.madukaife@unn.edu.ng

\begin{abstract}
\noindent In this research work, a total of 45 different estimators of the Shannon
differential entropy were reviewed. The estimators were mainly based
on three classes, namely: window size spacings ($m$), kernel density
estimation (KDE) and $k-$nearest neighbour ($k$NN) estimation. A
total of 16, 5 and 6 estimators were selected from each of the classes,
respectively, for comparison. The performances of the 27 selected estimators,
in terms of their bias values and root mean squared errors (RMSEs)
as well as their asymptotic behaviours, were compared through extensive
Monte Carlo simulations. The empirical comparisons were carried out
at different sample sizes of $n$ = 10, 50, and 100 and different
variable dimensions of $d$ = 1, 2, 3, and 5, for three groups of
continuous distributions according to their symmetry and support.
The results showed that the spacings based estimators generally performed
better than the estimators from the other two classes at $d$ = 1,
but suffered from non existence at $d\geq2$. The $k$NN based estimators
were generally inferior to the estimators from the other two classes
considered but showed an advantage of existence for all $d\geq1$.
Also, a new class of optimal window size was obtained and sets of
estimators were recommended for different groups of distributions
at different variable dimensions. Finally, the asymptotic biases,
variances and distributions of the 'best estimators' were considered. 


\end{abstract}

\section{Introduction}

The entropy of a random variable is a function which attempts to characterize
the randomness, unpredictability or uncertainty of the random variable.
It provides a measure of the average amount of information needed
to represent an event of a random variable from its probability distribution.
It can as well be said to be a means of quantifying the amount of
information in a random variable. Suppose $\boldsymbol{X}$ is a $d-$dimensional
finite discrete random vector ($d\geq1$) with sample space $\varTheta$
and probability function $P(\boldsymbol{X}=\boldsymbol{x})$. The
entropy of $\boldsymbol{X}$, denoted by $H(\boldsymbol{X})$, is
defined by Shannon (1948) as:
\begin{equation}
H(\boldsymbol{X})\thinspace=\,E\left[-log\left\{ P(\boldsymbol{X}=\boldsymbol{x})\right\} \right]\thinspace=\thinspace-\sum_{\boldsymbol{x}\in\varTheta}P(\boldsymbol{X}=\boldsymbol{x})log\left\{ P(\boldsymbol{X}=\boldsymbol{x})\right\} 
\end{equation}
 If the random variable $\boldsymbol{X}$ is continuous with distribution
function $F(\boldsymbol{x})$ and probability density function
$f(\boldsymbol{x})$, the Shannon entropy is usually regarded as differential
entropy with a functional form:
\begin{equation}
H(\boldsymbol{X})\thinspace=\,E\left[-log\left\{ f(\boldsymbol{x})\right\} \right]\thinspace=\thinspace-\int_{\mathbb{R}^{d}}f(\boldsymbol{x})log\left\{ f(\boldsymbol{x})\right\} d\boldsymbol{x}
\end{equation}

From the point of view of the estimation of expected values, a natural
estimator of $H(\boldsymbol{X})$ in (2) can be given by:
\begin{equation}
H_{n}(\boldsymbol{X})\thinspace=\thinspace-\frac{1}{n}\sum_{i=1}^{n}log\left\{ f_{n}(\boldsymbol{x}_{i})\right\} 
\end{equation}
where $f_{n}(\boldsymbol{x}_{i})$ is an estimated probability density
function, which can be obtained by either parametric or nonparametric
density estimation from a random sample of size $n$ independent observations.
If the distribution from where the random sample is drawn is known,
$f_{n}(\boldsymbol{x}_{i})$ is obtained parametrically as the pdf
with the parameter(s) substituted for by their unbiased sample estimate(s).
In real life however, the distributions from where random samples
are obtained are usually not known. In view of this, one may be left
with the option of using any of the nonparametric techniques, such
as kernel density estimation, $k-$nearest neighbour estimation or
the spacing estimation, to obtain $f_{n}(\boldsymbol{x}_{i})$ in
(3) which gives rise to a plug--in estimator of the Shannon entropy.
Unfortunately, the spacing technique exists for this purpose only
at $d=1$ and the plug--in estimators of Shannon entropy has continued
to receive criticisms. Estimation of $H(\boldsymbol{X})$ both in
(1) and (2) has therefore attracted the attention of several authors.
In fact, there are more than two scores of different estimators of
the Shannon entropy in the literature. Such estimators include Hutcheson
and Shenton (1974), and Vatutin and Michailov (1995), for the discrete
random variables in (1) as well as Vasicek (1976), Kozachenko and
Leonenko (1987), Ebrahimi et al. (1994), Wieczorkowski and Grzegorzewski
(1999), Yokota and Shiga (2004), Alizadeh Noughabi (2010), Park (2012),
Bouzebda et al. (2013), Alizadeh Noughabi and Alizadeh Noughabi (2013),
Hino et al. (2015), Lombardi and Pant (2016) and Kohansal and Rezakhah
(2016) to mention but a few, for the differential entropy in (2).

All those estimators are obtained using different nonparametric approaches,
thereby building classes of estimators where a class comprises all
the estimators obtained using a particular technique. Such techniques
that have been used include the window size ($m$) spacing, kernel
density estimation (KDE), $k-$th nearest neighbour ($k$NN) technique,
histogram density estimation as well as simple linear regression.
Amongst all these classes of estimators of the Shannon entropy, three
have remained dominant in the literature, namely: window size spacing,
kernel density estimation and the nearest neighbour. The purpose of
this work is to carryout an extensive review of the estimators. Feutrill
and Roughan (2021) had reviewed the Shannon entropy estimators. They
concluded by saying that there are still gaps, in particular, with
the development of improved non-parametric estimators for the differential
entropy rate as well as development of more efficient techniques for
non-parametric estimation of the Shannon entropy rate. Our review
in this work is carried out from a different perspective. The quality
of the estimators based on window size spacing is dependent on the
magnitude of the spacing $(m)$. In a similar manner, the quality
of the estimators based on KDE is dependent on the bandwidth size
$(h)$ while the quality of the estimators based on $k$NN is dependent
on the number of nearest neighbour $(k)$. For each of these techniques,
the present work seeks to obtain the best $m$, $h$, and $k$ respectively
that balances the bias--variance trade--off that is often encountered
in statistical estimation procedures.

The scope of the work embodied here is for random processes in a $d-$dimensional
space, $d\geq1$. Unfortunately, the spacing technique exists for
one--dimensional space only. The estimators under the class are compared
with each other as well as with those from other classes at the available
one--dimensional space while the rest are compared at different dimensions
including their asymptotic behaviours.

\section{Description of the Estimators}

In this section, all the competing estimators of the differential
entropy are classified into three, namely: window size $(m)$ spacings, KDE and $k$NN estimators. They are presented in what follows.

\subsection{Estimators based on sample spacings}

Suppose $x_{1},\thinspace x_{2},\thinspace.\thinspace.\thinspace.,\thinspace x_{n}$
is a random sample of size $n$ obtained from a random variable $X$
whose distribution function is $F(x)$ with probability density function
$f(x)$. The sample order statistics obtained from the random sample
is given as $X_{(1)},\thinspace X_{(2)},\thinspace.\thinspace.\thinspace.,\thinspace X_{(n)}$,
where $X_{(1)}\leq\thinspace X_{(2)}\leq\thinspace.\thinspace.\thinspace.\leq\thinspace X_{(n)}$.
Let $m$ be a non-negative integar. Then, $m$ sample spacing is defined
on $i$th order statistic by $X_{(i+m)}-X_{(i-m)}$, where $m\thinspace=\thinspace0,\thinspace1,\thinspace2,\thinspace.\thinspace.\thinspace.,\thinspace\frac{n}{2}$.
There are no fewer than 20 known estimators in this class. Their development is hinged on two basic mathematical
operations on (2) for one dimensional case, namely: a simple differential
transformation, and replacement of differential operator with a difference
operator on an empirical distribution function. It is possible to
transform (2) to the form:
\begin{equation}
H(X)\thinspace=\thinspace\int_{0}^{1}log\left\{ \frac{d}{dp}F^{-1}(p)\right\} dp
\end{equation}
where $p\in(0,1)$ and $F^{-1}(p)$ is the quantile function of $X$
such that the quantile function can be estimated by the empirical
quantile function, $F_{n}^{-1}(p)$; $\frac{(i-1)}{n}\thinspace<\thinspace p\thinspace<\thinspace\frac{i}{n}$,
$i\thinspace=\thinspace1,\thinspace2,\thinspace.\thinspace.\thinspace.,\thinspace n$.
Using this procedure, Vasicek (1976), van Es (1992) including up till
about Madukaife (2023) have obtained different estimators of
the Shannon differential entropy. Again, they are presented in what follows.

Vasicek (1976) introduced one of the earliest estimators of the differential
entropy by transforming the Shannon entropy in (2) to the form in
(4) where the author estimated (4) by replacing the quantile function
with the empirical quantile function and using a difference operator
in place of the differential operator to obtain the estimated derivative
of $F^{-1}(p)$ in (4) as $\left[X_{(i+m)}-X_{(i-m)}\right]\frac{n}{2m}$
for $\frac{(i-1)}{n}\thinspace<p<\thinspace\frac{i}{n}$; $i\thinspace=\thinspace m+1.\thinspace m+2,\thinspace.\thinspace.\thinspace.,\thinspace n-m$.
With this and some other conditions on the estimated derivative, the
Vasicek (1976) estimator of the $H(X)$ is obtained as:
\[
HV_{m,n}\thinspace=\thinspace\frac{1}{n}\sum_{i=1}^{n}log\left\{ \frac{n}{2m}\left[X_{(i+m)}-X_{(i-m)}\right]\right\} ;
\]
where $X_{(i)}\thinspace=\thinspace X_{(1)}$ for $i<1$ and $X_{(i)}\thinspace=\thinspace X_{(n)}$
for $i>n$. The work proved that the convergence of the estimator
is such that $\left[HV_{m,n}-H(X)\right]\thinspace\stackrel{P}{\rightarrow}\thinspace E_{m,n}$
as $n\thinspace\rightarrow\thinspace\infty,\thinspace m\thinspace\rightarrow\thinspace\infty$
and $m/n\thinspace\rightarrow\thinspace0$, where $E_{m,n}\thinspace=\thinspace log(n)-log(2m)+\left(1-\frac{2m}{n}\right)\varPsi(2m)-\varPsi(n+1)+\frac{2}{n}\sum_{i=1}^{m}\varPsi(i+m-1)$
and $\varPsi(x)\thinspace=\thinspace\frac{d}{dx}log\varGamma(x)\thinspace=\thinspace\frac{\varGamma^{\prime}(x)}{\varGamma(x)}$
is the digamma function, see Abramowitz and Stegun (1972).

Dudewicz and Van der Meulen (1987) studied the entropy estimator due
to Vasicek (1976). Upon generalization, they obtained an estimator
of the Shannon entropy defined by: 
\[
HD_{m,n}\thinspace=\thinspace-\frac{1}{n}\sum_{i=2-m}^{0}\sum_{j=1}^{i+m-1}\left(\frac{X_{(i+m)}-X_{(i+m-1)}}{X_{(j+m)}-X_{(j-m)}}\right)log\left\{ \frac{2}{n}\sum_{j=1}^{i+m-1}\left(\frac{1}{X_{(j+m)}-X_{(j-m)}}\right)\right\} 
\]
\[
-\frac{1}{n}\sum_{i=1}^{n+1-m}\sum_{j=i}^{i+m-1}\left(\frac{X_{(i)}+X_{(i+m)}-X_{(i-1)}-X_{(i+m-1)}}{X_{(j+m)}-X_{(j-m)}}\right)log\left\{ \frac{2}{n}\sum_{j=1}^{i+m-1}\left(\frac{1}{X_{(j+m)}-X_{(j-m)}}\right)\right\} 
\]
\[
-\frac{1}{n}\sum_{i=n+2-m}^{n}\sum_{j=1}^{n}\left(\frac{X_{(i)}-X_{(i-1)}}{X_{(j+m)}-X_{(j-m)}}\right)log\left\{ \frac{2}{n}\sum_{j=1}^{n}\left(\frac{1}{X_{(j+m)}-X_{(j-m)}}\right)\right\} .
\]

van Es (1992) introduced the problem of estimation of a general phi
functional of the form 
\begin{equation}
T(f)\thinspace=\thinspace\int_{-\infty}^{\infty}f(x)\Phi\left(f(x)\right)w(x)dx
\end{equation}
 where $f$ and $w$ are real valued functions on $[0,\thinspace\infty)$
satisfying certain regularity conditions. By substituting $-log\left(f(x)\right)$
for $\Phi\left(f(x)\right)w(x)$ in (5) and estimating independently
via spacing, the bias corrected estimator due to van Es (1992) is
obtained by:
\[
HE_{m,n}\thinspace=\thinspace\frac{1}{n-m}\sum_{i=1}^{n-m}log\left\{ \frac{n+1}{m}\left(X_{(i+m)}-X_{(i)}\right)\right\} \thinspace+\thinspace\sum_{k=m}^{n}\frac{1}{k}\thinspace+\thinspace log(m)-log(n+1)
\]
where $k\thinspace=\thinspace m,\thinspace m+1,\thinspace.\thinspace.\thinspace.,\thinspace n$.
The work showed that for a fixed $m$, $HE_{m,n}\thinspace\stackrel{a.s}{\rightarrow}\thinspace H(X)$
as $m,n\thinspace\rightarrow\thinspace\infty$ and derived the asymptotic
null distribution of the estimator as $n^{1/2}\left[HE_{m,n}\thinspace-\thinspace H(X)\right]\thinspace\stackrel{D}{\rightarrow}\thinspace N\left(0,\thinspace var\left(log(f(X_{1})\right)\right)$
which proves it as a consistent estimator.

Ebrahimi et al. (1994) obtained two estimators of the differential
entropy. They argued that the slope at each sample point, $x_{i}$,
$s_{i}(m,n)\thinspace=\thinspace\frac{n}{2m}\left(X_{(i+m)}-X_{(i-m)}\right)$
obtained by Vasicek (1976) upon which he obtained his estimator is
incorrect for all $i\thinspace\leq\thinspace m$ and all $i\thinspace\geq\thinspace n-m+1$.
As a result, they modified the slope (estimated derivative of $F^{-1}(p)$)
due to Vasicek (1976) to obtain a new estimator given by:
\[
HE_{m,n}^{(1)}\thinspace=\thinspace\frac{1}{n}\sum_{i=1}^{n}log\left\{ \frac{n}{c_{i}m}\left(X_{(i+m)}-X_{(i-m)}\right)\right\} ,
\]
where $c_{i}\thinspace=\thinspace\begin{cases}
\begin{array}{c}
1+\frac{i-1}{m},\thinspace\thinspace\thinspace\thinspace\thinspace\thinspace\thinspace\thinspace\thinspace\thinspace\thinspace\thinspace1\leq i\leq m,\\
2,\,\thinspace\thinspace\thinspace\thinspace\thinspace\thinspace\thinspace\thinspace\,m+1\leq i\leq n-m,\\
1+\frac{n-i}{m},\thinspace n-m+1\leq i\leq n,
\end{array}\end{cases}$ $X_{(i-m)}\thinspace=\thinspace X_{(1)}$ for $i\leq m$ and $X_{(i+m)}\thinspace=\thinspace X_{(n)}$
for $i\geq n-m$ and they derived that $HE_{m,n}^{(1)}\thinspace=\thinspace HV_{m,n}\thinspace+\thinspace\frac{2}{n}\left[mlog(2m)+log\left\{ \frac{(m-1)!}{(2m-1)!}\right\} \right]$,
where $HV_{m,n}$ is the Vasicek (1976) estimator.

Again, by yet another modification of the estimated derivative of
$F^{-1}(p)$ due to Vasicek (1976), they obtained another estimator
given by:
\[
HE_{m,n}^{(2)}\thinspace=\thinspace\frac{1}{n}\sum_{i=1}^{n}log\left\{ \frac{n}{d_{i}m}\left(Z_{(i+m)}-Z_{(i-m)}\right)\right\} ,
\]
where $d_{i}\thinspace=\thinspace\begin{cases}
\begin{array}{c}
1+\frac{i+1}{m}-\frac{i}{m^{2}},\thinspace\thinspace\thinspace\thinspace\thinspace\thinspace\thinspace\thinspace\thinspace\thinspace\thinspace\thinspace1\leq i\leq m,\\
2,\,\thinspace\thinspace\thinspace\thinspace\thinspace\thinspace\thinspace\thinspace\,m+1\leq i\leq n-m-1,\\
1+\frac{n-i}{m+1},\thinspace\thinspace\thinspace\thinspace\thinspace\thinspace\thinspace\thinspace\thinspace\thinspace\thinspace\,n-m\leq i\leq n,
\end{array}\end{cases}$ $Z_{(i-m)}\thinspace=\thinspace a+\frac{i-1}{m}\left(X_{(1)}-a\right)\thinspace=\thinspace X_{(1)}-\frac{m-i+1}{m}\left(X_{(1)}-a\right);\thinspace1\leq i\leq m$,
$Z_{(i)}\thinspace=\thinspace X_{(i)};\thinspace m+1\leq i\leq n-m-1$,
$Z_{(i+m)}\thinspace=\thinspace b-\frac{n-i}{m}\left(b-X_{(n)}\right)\thinspace=\thinspace X_{(n)}+\frac{m+i-n}{m}\left(b-X_{(n)}\right);\thinspace n-m\leq i\leq n$,
and $a,b$ are two known constants such that $P\left(a\leq X\leq b\right)\thinspace\simeq\thinspace1$.
They concluded that $HE_{m,n}^{(1)}\thinspace>\thinspace HV_{m,n}$,
$HE_{m,n}^{(2)}\thinspace>\thinspace HV_{m,n}$ and as a result, that
the compuation of sample entropy based on $HE_{m,n}^{(1)}$ and $HE_{m,n}^{(2)}$
is more in line with the maximum entropy principle than the $HV_{m,n}$
estimator due to Vasicek (1976). Also, they proved the convergence
in probability of both $HE_{m,n}^{(1)}$ and $HE_{m,n}^{(2)}$ to
the $H(X)$ as $n\thinspace\rightarrow\thinspace\infty,\thinspace m\thinspace\rightarrow\thinspace\infty$
and $m/n\thinspace\rightarrow\thinspace0$. Through empirical studies,
they showed that $HE_{m,n}^{(2)}$ is more efficient than the $HE_{m,n}^{(1)}$.

Correa (1995) obtained the entropy of a random variable $X$ in (4)
with the estimated derivative of $F^{-1}(p)$ to be in the form:
\begin{equation}
H(X)\thinspace=\thinspace-\frac{1}{n}\sum_{i=1}^{n}log\left\{ \frac{\frac{i+m}{n}-\frac{i-m}{n}}{X_{(i+m)}-X_{(i-m)}}\right\} .
\end{equation}
With (6), the density of $F(x)$ was estimated in the interval $\left(X_{(i-m)},\thinspace X_{(i+m)}\right)$
using a local linear regression model based on $2m+1$ points. This
resulted in the estimator of the differential entropy due to Correa
(1995). The estimator is given by:
\[
HC_{m,n}\thinspace=\thinspace-\frac{1}{n}\sum_{i=1}^{n}log\left(b_{i}\right),
\]
where $b_{i}\thinspace=\thinspace\frac{\sum_{j=i-m}^{i+m}\left(X_{(j)}-\overline{X}_{(j)}\right)\left(\frac{j}{n}-\frac{i}{n}\right)}{\sum_{j=i-m}^{i+m}\left(X_{(j)}-\overline{X}_{(i)}\right)^{2}}$,
$\overline{X}_{(i)}\thinspace=\thinspace\sum_{j=i-m}^{i+m}\frac{X_{(j)}}{2m+1}$,
$X_{(i)}\thinspace=\thinspace X_{(1)}$ if $i\,<\thinspace1$ and
$X_{(i)}\thinspace=\thinspace X_{(n)}$ if $i\,>\thinspace n$. The
estimator was compared with the Vasicek estimator empirically and
it was shown to have smaller mean squared error (MSE) than the Vasicek
estimator.

Vasicek (1976) through the convergence tructure of his estimator opined
an idea for bias correction of the $HV_{m,n}$ estimator. Applying
this correction method, Wieczorkowski and Grzegorzewski (1999) obtained
a bias corrected estimator of the differential entropy. The estimator
is given by:
\[
HW_{m,n}^{(1)}\thinspace=\thinspace HV_{mn}\thinspace-\thinspace log\left(n\right)\thinspace+\thinspace log\left(2m\right)\thinspace-\thinspace\left(1-\frac{2m}{n}\right)\varPsi\left(2m\right)\thinspace+\thinspace\varPsi\left(n+1\right)\thinspace-\thinspace\frac{2}{n}\sum_{i=1}^{m}\varPsi\left(i+m-1\right),
\]
where $HV_{mn}$ is the entropy estimator due to Vasicek (1976), $\varPsi\left(x\right)$
is the digamma function defined for the random variable $X$ as $\varPsi\left(x\right)\thinspace=\thinspace\frac{d}{dx}\varGamma(x)\thinspace=\thinspace\frac{\varGamma^{\thinspace\prime}(x)}{\varGamma(x)}$.

Again, they obtained the Jacknife bias corrected estimator of the
Shannon entropy by Jacknifing the Correa (1995) differential entropy
estimator, $HC_{m,n}$. The resultant Jacknife estimator is given
by: 
\[
HW_{m,n}^{(2)}\thinspace=\thinspace HC_{m,n}\thinspace-\thinspace\left(n-1\right)\left(HC_{m,n}^{*}-HC_{m,n}\right),
\]
where $HC_{m,n}^{*}\thinspace=\thinspace n^{-1}\sum_{i=1}^{n}\left(HC_{m,n}\right)_{i}$.
Through extensive simulation studies, they compared their estimators
with those of Vasicek (1976), van Es (1992) and Correa (1995).

Pasha et al. (2005) studied the differential entropy estimators due
to Ebrahimi et al. (1994). As a result, they obtained a modification
of the $HE_{m,n}^{(2)}$ estimator as a new one and derived some characterizations
of the new modified estimator. It is given by:
\[
HP_{m,n}\thinspace=\thinspace\frac{1}{n}\sum_{i=1}^{n}log\left\{ \frac{n}{d_{i}m}\left(Z_{(i+m)}-Z_{(i-m)}\right)\right\} ,
\]
where $d_{i}\thinspace=\thinspace\begin{cases}
\begin{array}{c}
1+\frac{i}{i-1+m},\thinspace\thinspace\thinspace\thinspace\thinspace\thinspace\thinspace\thinspace\thinspace\thinspace\thinspace\thinspace1\leq i\leq m,\\
2,\,\thinspace\thinspace\thinspace\thinspace\thinspace\thinspace\thinspace\thinspace\thinspace\thinspace\thinspace\thinspace\,m+1\leq i\leq n-m,\\
1+\frac{n-i+1}{n-i+m},\thinspace\,n-m+1\leq i\leq n,
\end{array}\end{cases}$ if $m\thinspace<\thinspace\frac{n}{2}$, $d_{i}\thinspace=\thinspace\begin{cases}
\begin{array}{c}
1+\frac{i}{i-1+m},\thinspace\thinspace\thinspace\thinspace\thinspace\thinspace\thinspace\thinspace\thinspace\thinspace\thinspace\thinspace1\leq i\leq m,\\
1+\frac{n-i+1}{n-i+m},\thinspace\,n-m+1\leq i\leq n,
\end{array}\end{cases}$ if $m\thinspace=\thinspace\frac{n}{2}$, and $Z_{(i)}$ is as defined
by Ebrahimi et al. (1994) with the bounds of support, $a$ and $b$,
estimated by $a_{n}\thinspace=\thinspace X_{(1)}-\frac{X_{(n)}-X_{(1)}}{n-1}$
and $b_{n}\thinspace=\thinspace X_{(n)}-\frac{X_{(n)}-X_{(1)}}{n-1}$
respectively. They proved the convergence of their estimator, in probability,
as $n,m\thinspace\rightarrow\thinspace\infty$, $m/n\thinspace\rightarrow\thinspace0$
as well as the asymptotic unbiasedness of the estimator (ie $E(HP_{m.n})\thinspace\rightarrow\thinspace H(X)$
as $n,m\thinspace\rightarrow\thinspace\infty$, $m/n\thinspace\rightarrow\thinspace0$).

On their own, Alizadeh Noughabi and Arghami (2010) studied the $HE_{m,n}^{(1)}$
estimator of the differential entropy due to Ebrahimi et al. (1994)
and showed that the estimator underestimated the Shannon entropy at
small sample sizes in almost all cases of distributional laws. They
therefore modified the coefficients of the estimator to obtain a new
estimator which is given by:
\[
HA_{m,n}\thinspace=\thinspace\frac{1}{n}\sum_{i=1}^{n}log\left\{ \frac{n}{a_{i}m}\left[X_{(i+m)}-X_{(i-m)}\right]\right\} ,
\]
where $a_{i}\thinspace=\thinspace\begin{cases}
\begin{array}{c}
1,\thinspace\thinspace\thinspace\thinspace\thinspace\thinspace\thinspace\thinspace\thinspace\thinspace\thinspace\thinspace1\leq i\leq m\\
2,\thinspace m+1\leq i\leq n-m\\
1,\thinspace n-m+1\leq i\leq n
\end{array}\end{cases}$; $X_{(i-m)}\thinspace=\thinspace X_{(1)}$ for $i\leq m$ and $X_{(i+m)}\thinspace=\thinspace X_{(n)}$
for $i\geq n-m$. Again, they proved the invariance property on the
efficiency of their estimator with respect to changes in the scale
of the random observations and established among other things that
$HA_{m,n}\thinspace>\thinspace HV_{m,n}$; $HA_{m,n}\thinspace\geq\thinspace HE{}_{m,n}^{(2)}$
and that $HA_{m,n}\thinspace\rightarrow H(X)$ as $n,m\thinspace\rightarrow\thinspace\infty$,
$m/n\thinspace\rightarrow\thinspace0$.

In a yet another modification of the Vasicek (1976) estimator based
on correction of coefficients but different from all the other previous
modifications, Zamanzade and Arghami (2011) insisted that $\left(n/2m\right)\left(X_{(i+m)}-X_{(i-m)}\right)$
is not a good approximation for the slope when $i\leq m$ or $i\geq n-m+1$,
which was the standpoint of Ebrahimi et al. (1994) and others which
led to their proposed estimators. They suggested that replacing the
$c_{i}'s$ in Ebrahimi et al. (1994) for $1\leq i\leq m$ and $n-m+1\leq i\leq n$
with smaller numbers could reduce the bias without affecting the standard
deviation of the estimator. Based on this, they proposed a bias-corrected
estimator of the Shannon entropy, which is given by: 
\[
HZ_{m,n}\thinspace=\thinspace\frac{1}{n}\sum_{i=1}^{n}log\left\{ \frac{n}{a_{i}m}\left(X_{(i+m)}-X_{(i-m)}\right)\right\} 
\]
where $a_{i}\thinspace=\thinspace\begin{cases}
\begin{array}{c}
\frac{i}{m},\,\,\,\,\,\,\,\,\,\,\,\,\,\,\,1\leq i\leq m\\
2,\,\,\,m+1\leq i\leq n-m\\
\frac{n-i+1}{m},\,\,n-m+1\leq i\leq n
\end{array}\end{cases}.$ In addition to stating that the efficiency of the $HZ_{m,n}$ is
invariant with respect to changes in the scale of the original observations
as well as the convergence in probability of the estimator as $n,m\thinspace\rightarrow\thinspace\infty$,
$m/n\thinspace\rightarrow\thinspace0$, they also stated that the
optimal $m$ suggested by Grzegorzewski and Wieczorkowski (1999) ($m\thinspace=\thinspace[\sqrt{n}\thinspace+0.5]$)
is slightly not correct and suggested an optimal $m$ to be such that
$\begin{cases}
\begin{array}{c}
\left[\sqrt{n}+0.5\right]\thinspace\thinspace\thinspace\thinspace\thinspace\mathrm{if}\thinspace n\thinspace\leq\thinspace8\\
\left[\sqrt{n}+0.5\right]-1\thinspace\mathrm{if}\thinspace n\thinspace\geq9
\end{array}\end{cases}$.

Park and Shin (2012) criticized the Vasicek (1976) estimator of the
Shannon entropy as well as all its modifications in the literature
as having a defectiveness caused by replacing $X_{(i)}s$ for $i<1$
and $i>n$ with $X_{(1)}$ and $X_{(n)}$ respectively. They employed
the nonparametric density estimator of Park and Park (2003) to obtain
an estimator of the entropy defined by:
\[
HP_{m,n}\thinspace=\thinspace\frac{1}{n}\sum_{i=1}^{n}log\left\{ n\left(\xi_{i+1}^{\prime}-\xi_{i}^{\prime}\right)\right\} 
\]
where $\xi_{i}^{\prime}\thinspace=\thinspace\frac{\left(X_{(i-m)}+X_{(i-m+1)}+\thinspace.\thinspace.\thinspace.\thinspace+X_{(i+m-1)}\right)}{2m}$.

Al-Omari (2014) introduced the concept of sampling scheme with which
the sample whose entropy measure is required to the approprateness
of entropy estimators and the estimators are intended to improving
the Vasicek (1976) since its $s_{i}\left(m,n\right)\thinspace=\thinspace\frac{n}{2m}\left(X_{(i+m)}-X_{(i-m)}\right)$
is not a good formula for the slope when $i\leq m$ or $i\geq n-m+1$.
The modifications to the slope gave rise to three estimators depending
on whether the sample is obtained through simple random sampling (SRS),
ranked set sampling (RSS) or double ranked set sampling (DRSS) techniques.
Since most samples in nature are obtained by SRS, only the estimator
for samples obtained by SRS is considered in this study. It is defined
by:
\[
HA_{m,n}^{(1)}\thinspace=\thinspace\frac{1}{n}\sum_{i=1}^{n}log\left\{ \frac{n}{c_{i}m}\left(X_{(i+m)}-X_{(i-m)}\right)\right\} 
\]
where $c_{i}\thinspace=\thinspace\begin{cases}
\begin{array}{c}
1+\frac{1}{2},\,\,\,\,\,\,\,\,\,\,\,\,\,\,\,1\leq i\leq m\\
2,\,\,\,m+1\leq i\leq n-m\\
1+\frac{1}{2},\,\,n-m+1\leq i\leq n
\end{array}\end{cases}$; $X_{(i-m)}\thinspace=\thinspace X_{(1)}$ for $i\leq m$ and $X_{(i+m)}\thinspace=\thinspace X_{(n)}$
for $i\geq n-m$. The author equally showed that $HA_{m,n}^{(1)}$ is related
to $HV_{m,n}$ by:
\[
HA_{m,n}^{(1)}\thinspace=\thinspace HV_{m,n}\thinspace+\thinspace\frac{2}{n}mlog\left\{ \frac{4}{3}\right\} .
\]

Al-Omari (2015) further modified the $HA_{m,n}^{(1)}$ estimator of
Al-Omari (2014) which is based on a SRS technique. This further modification
resulted in a new estimator given by:
\[
HA_{m,n}^{(2)}\thinspace=\thinspace\frac{1}{n}\sum_{i=1}^{n}log\left\{ \frac{n}{\varepsilon_{i}m}\left(X_{(i+m)}-X_{(i-m)}\right)\right\} 
\]
where $\varepsilon_{i}\thinspace=\thinspace\begin{cases}
\begin{array}{c}
1+\frac{1}{4},\,\,\,\,\,\,\,\,\,\,\,\,\,\,\,1\leq i\leq m\\
2,\,\,\,m+1\leq i\leq n-m\\
1+\frac{1}{4},\,\,n-m+1\leq i\leq n
\end{array}\end{cases}$; $X_{(i-m)}\thinspace=\thinspace X_{(1)}$ for $i\leq m$ and $X_{(i+m)}\thinspace=\thinspace X_{(n)}$
for $i\geq n-m$. Also, the estimator $HA_{m,n}^{(2)}$ is related
to the Vasicek (1976) estimator by:
\[
HA_{m,n}^{(2)}\thinspace=\thinspace HV_{mn}\thinspace+\thinspace\frac{2m}{n}mlog\left\{ \frac{8}{5}\right\} .
\]
 He showed that both $HA_{m,n}^{(1)}$ and $HA_{m,n}^{(2)}$ converge
in probability to $H(X)$ as $m,n\thinspace\rightarrow\thinspace\infty,$
$m/n\thinspace\rightarrow\thinspace0$.

Kohansal and Rezakhah (2016) studied the $HE_{m,n}$ and $HW_{m,n}$
estimators of van Es (1992) and Wieczorkowski and Grzegorzewski (1999)
respectively and adapted the two to theirs by first smoothening the
sample data using the moving average (MA) method. Hence, they obtained:
\[
HK_{m,n}^{(1)}\thinspace=\thinspace\frac{1}{n}\sum_{i=1}^{n}log\left\{ Y_{(i+m-1)}-Y_{(i-m-1)}\right\} \thinspace+\thinspace c
\]
and 
\[
HK_{m,n}^{(2)}\thinspace=\thinspace\frac{1}{n-m}\sum_{i=2}^{n-m}log\left\{ Y_{(i+m-1)}-Y_{(i-1)}\right\} \thinspace+\thinspace\sum_{k=m}^{n}\frac{1}{k}
\]
where $Y_{(i)}\thinspace=\thinspace\begin{cases}
\begin{array}{c}
\frac{\sum_{j=1}^{i}X_{(j)}}{i},\thinspace\thinspace\thinspace\thinspace\thinspace\thinspace\thinspace\thinspace\thinspace\thinspace\thinspace\thinspace\thinspace\thinspace\thinspace\thinspace\thinspace\thinspace\thinspace\thinspace\thinspace\thinspace\thinspace\thinspace\thinspace\thinspace\thinspace\thinspace\thinspace\thinspace\thinspace\thinspace\thinspace\thinspace\thinspace1\leq i\leq\frac{(w-1)}{2}\\
\frac{\sum_{j=i-(w-1)/2}^{i+(w-1)/2}X_{(j)}}{w},\thinspace\frac{(w+1)}{2}\leq i\leq n-\frac{(w-1)}{2};\thinspace i\thinspace=\thinspace1,\thinspace2,\thinspace.\thinspace.\thinspace.,\thinspace n\\
\frac{\sum_{j=i}^{n}X_{(j)}}{n-i+1},\thinspace\thinspace\thinspace\thinspace\thinspace\thinspace\thinspace\thinspace\thinspace\thinspace\thinspace\thinspace\thinspace\thinspace\thinspace\thinspace\thinspace\thinspace\thinspace\thinspace\thinspace\thinspace\thinspace\thinspace\thinspace\thinspace\thinspace\thinspace\thinspace\thinspace\thinspace\thinspace\thinspace\thinspace\thinspace n-\frac{(w-3)}{2}\leq i\leq n
\end{array}\end{cases}$ and $c\thinspace=\thinspace-\left(1-\frac{2m}{n}\right)\psi\left(2m\right)+\psi\left(n+1\right)-\frac{2}{n}\sum_{i=1}^{m}\psi\left(i+m-1\right)$.
They however did not give any condition for $w$ but showed that the
scale of the random variable has no effect on the accuracy of the
two estimators and proved that if $m\thinspace=\thinspace o(n)$,
then $HK_{m,n}^{(2)}\thinspace\stackrel{Pr}{\rightarrow}\thinspace H(X)$
as $n\thinspace\rightarrow\thinspace\infty$.

In a quest to obtain a consistently good estimator for the entropy
of a continuous random variable, Al-Omari (2016) further again modified
the Al-Omari (2014) estimator to have:
\[
HA_{m,n}^{(3)}\thinspace=\thinspace\frac{1}{n}\sum_{i=1}^{n}log\left\{ \frac{n}{\nu_{i}m}\left(X_{(i+m)}-X_{(i-m)}\right)\right\} 
\]
where $\nu_{i}\thinspace=\thinspace\begin{cases}
\begin{array}{c}
1+\frac{i-1}{m},\,\,\,\,\,\,\,\,\,\,\,\,\,\,\,1\leq i\leq m\\
2,\,\,\,m+1\leq i\leq n-m\\
1+\frac{n-i}{2m},\,\,n-m+1\leq i\leq n
\end{array}\end{cases}$; $X_{(i-m)}\thinspace=\thinspace X_{(1)}$ for $i\leq m$ and $X_{(i+m)}\thinspace=\thinspace X_{(n)}$
for $i\geq n-m$. Also, like in Al-Omari (2014), he obtained the relationship between
the estimator $HA_{m,n}^{(3)}$ and the Vasicek (1976) estimator as
given by $HA_{m,n}^{(3)}\thinspace=\thinspace HV_{m,n}\thinspace+\thinspace\frac{1}{n}\sum_{i=1}^{n}log\left\{ \frac{2}{\nu_{i}}\right\} $
and proved its convergence in probability to $H(X)$ as $m,n\thinspace\rightarrow\thinspace\infty$
and $m/n\thinspace\rightarrow\thinspace0$.

Bitaraf et al. (2017) replaced the distribution function $F(x)$ with
the empirical distribution function $F_{n}(x)$ to obtain
\begin{equation}
\frac{d}{dp}F^{-1}\left(p\right)\thinspace\simeq\thinspace T_{i}
\end{equation}
where $T_{i}\thinspace=\thinspace\frac{T_{i0}+T_{i1}}{2};\thinspace T_{ij}\thinspace=\thinspace\frac{n\left(X_{(i+m-j)}-X_{(i-m+j)}\right)}{w_{j}(m-j)};\thinspace j\thinspace=\thinspace0,\thinspace1$
for $\frac{(i-1)}{n}\thinspace\leq p\thinspace\leq\frac{i}{n}$ and
$i\thinspace=\thinspace m+1,\thinspace m+2,\thinspace.\thinspace.\thinspace.,\thinspace n-m$
such that 

$w_{ij}\thinspace=\thinspace\begin{cases}
\begin{array}{c}
1,\thinspace\thinspace\thinspace\thinspace\thinspace\thinspace\thinspace\thinspace\thinspace\thinspace\thinspace\thinspace1\leq i\leq m-j\\
2,\thinspace m-j+1\leq i\leq n-m+j\\
1,\thinspace n-m+j+1\leq i\leq n
\end{array}\end{cases}$. With the results of (7), they defined two estimators of the Shannon entropy
measure as:
\[
HB_{m,n}^{(1)}\thinspace=\thinspace\frac{1}{n}\sum_{i=1}^{n}log\left\{ T_{i}\right\} 
\]
and 
\[
HB_{m,n}^{(2)}\thinspace=\thinspace\frac{1}{2n}\left[log\left(T_{1}\right)+log\left(T_{n}\right)+2\sum_{i=2}^{n-1}log\left(T_{i}\right)\right].
\]
 They proved that both $HB_{m,n}^{(1)}$ and $HB_{m,n}^{(2)}$ converge
in probability to $H(X)$ as $m,n\thinspace\rightarrow\thinspace\infty,\thinspace m/n\thinspace\rightarrow\thinspace0$.
Also, they established that the efficiency of the two estimators is
not affected by any change in the scale of the observations.

Still leveraging on the fact that the slope, $\left(n/2m\right)\left(X_{(i+m)}-X_{(i-m)}\right)$
due to Vasicek (1976) is not correct when $i\leq m$ or $i\geq n-m+1$,
Alizadeh Noughabi and Jarrahiferiz (2019) proposed a new slope corrected
estimator. It is given by:
\[
HJ_{m,n}\thinspace=\thinspace-\frac{1}{n}\sum_{i=1}^{n}log\left\{ S_{i}(n,m)\right\} 
\]
where $\thinspace S_{i}(m,n)\thinspace=\thinspace\begin{cases}
\begin{array}{c}
\frac{m/n}{X_{(i+m)}-X_{(1)}},\thinspace\thinspace\thinspace\thinspace\thinspace\thinspace\thinspace\thinspace\thinspace\thinspace\thinspace\thinspace\thinspace\thinspace\thinspace\thinspace\thinspace\thinspace\thinspace\thinspace\thinspace\thinspace\,\thinspace\thinspace\thinspace\thinspace\thinspace\thinspace1\leq j\leq m\\
\frac{\sum_{j=i-m}^{i+m}(X_{(j)}-\overline{X}_{(i)})(j-i)}{n\sum_{j=i-m}^{i+m}(X_{(j)}-\overline{X}_{(i)})^{2}},\thinspace\thinspace\thinspace\thinspace\thinspace\thinspace\thinspace\thinspace\thinspace\thinspace\thinspace\thinspace\thinspace\thinspace\thinspace m+1\leq j\leq n-m\\
\frac{m/n}{X_{(n)}-X_{(i-m)}},\thinspace\thinspace\thinspace\thinspace\thinspace\thinspace\thinspace\thinspace\thinspace\thinspace\,\thinspace\thinspace\thinspace\thinspace\thinspace\thinspace\thinspace n-m+1\leq j\leq n
\end{array}\end{cases}.$

They went further to prove that the estimator is scale invariant and
also showed that the $HJ_{m,n}$ converges in probability to the $H(X)$
as $n\thinspace\rightarrow\thinspace\infty$.

Following from Bitaraf et al. (2017) estimation methodology, Madukaife
(2023) rescaled the internal $j$th spacing of the former to 2$j$
which gave rise to a new estimator. It is given by
\[
HM_{mn}\thinspace=\thinspace\frac{1}{n}\sum_{i=1}^{n}log\left\{ \frac{1}{2}\left(T_{i0}+T_{i1}\right)\right\} 
\]
where $T_{ij}\thinspace=\thinspace\frac{n}{w_{ij}(m-2j)}\left\{ X_{(i+m-j)}-X_{(i-m+j)}\right\} $
and $w_{ij}\thinspace=\thinspace\begin{cases}
\begin{array}{c}
1+\frac{1}{3},\thinspace\thinspace\thinspace\thinspace\thinspace\thinspace\thinspace\thinspace\thinspace\thinspace\thinspace\thinspace1\leq i\leq m-j\\
2,\thinspace m-j+1\leq i\leq n-m+j\\
1+\frac{1}{3},\thinspace n-m+j+1\leq i\leq n
\end{array}\end{cases}$; $X_{(i-m+j)}\thinspace=\thinspace X_{(1)}$ for $i\leq m-j$ and
$X_{(i+m-j)}\thinspace=\thinspace X_{(n)}$ for $i\geq n-m+j$; $j\thinspace=\thinspace0,1$.

Besides the estimators described under the present category, it has
equally been obtained for some special class of distributions. For
instance, Chaji and Zografos (2018) obtained the spacing--type estimator
of the Shannon differential entropy for beta--generated distributions.

\subsection{Estimators based on kernel density estimation}

Suppose $x_{1},\thinspace x_{2},\thinspace.\thinspace.\thinspace.,\thinspace x_{n}$
is a random sample of $n$ independent observations from a univariate
random variable $X$ whose distributional law with probability density
function (pdf) $f(x)$ is unknown. Rosenblatt (1956) pioneered a nonparametric
estimation of the pdf and Parzen (1962) obtained it in more concrete
terms (with its properties) as:
\begin{equation}
\hat{f}(x)\thinspace=\thinspace\frac{1}{nh}\sum_{i=1}^{n}K\left(\frac{x-x_{i}}{h}\right)
\end{equation}
where $K$ is the smooth function known as the kernel function, with
$h>0$ as the smoothing parameter known as the bandwidth. It is
the bandwith that controls the amount of smoothing in $K$ and a kernel
function $K(t)$ is such that:
\begin{equation}
\begin{cases}
\begin{array}{c}
\int_{-\infty}^{\infty}K(t)dt\thinspace=\thinspace1;\\
K(t)\thinspace=\thinspace K(-t)\thinspace\forall\thinspace t\in R;\\
\int_{-\infty}^{\infty}tK(t)dt\thinspace=\thinspace0;\\
\int_{-\infty}^{\infty}t^{2}K(t)dt\thinspace\coloneqq\thinspace k_{2}\thinspace<\thinspace\infty
\end{array}\end{cases}
\end{equation}
The estimator in (8) is known as the kernel density estimator (KDE)
and the estimation method is called kernel density estimation (KDE).
A good number of functions satisfying (9) have been mostly used in
the literature, such as the Epanechnikov, triangular, Gaussian (normal),
rectangular and cosine functions. Among all the kernel functions mostly
used, the Epanechnikov kernel is the most efficient (Wand \& Jones,
1995) with minimum mean integrated squared error (MISE) at a specified
bandwidth, $h$. Recently however, some nonsymmetric kernels have
been developed, see for instance Chen (2000), Scaillet (2001), Jin
and Kawczak (2003) as well as Weglarczyk (2018).

One of the earliest estimators of Shannon entropy in this group was
introduced by Dmitriev and Tarasenko (1974). They suggested a direct
plug-in of the KDE of $f(x)$ in (2) so as to obtain an integral estimate:
\begin{equation}
H_{n}(X)\thinspace=\thinspace-\int_{A_{h}}\hat{f}(x)log\left\{ \hat{f}(x)\right\} dx
\end{equation}
as an appropriate estimator of $H(X)$, where $\hat{f}(x)$ is the
kernel density estimator with bandwidth $h$. The estimation in (10)
however requires numerical integration, which is not in any way easy
so long as $\hat{f}(x)$ remains a KDE (Joe, 1989). As an alternative,
Ahmad and Lin (1976) rather, expressed entropy in the form $H(X)\thinspace=\thinspace-\int log\left\{ f(x)\right\} dF(x)$
and substituted the empirical distribution function $F_{n}(x)$ and
the kernel density function $\hat{f}(x)$ respectively for $F(x)$and
$f(x)$, to have 
\begin{equation}
H_{n}(X)\thinspace=\thinspace-\int log\left\{ \hat{f}(x)\right\} dF_{n}(x)\thinspace=\thinspace-\frac{1}{n}\sum_{i=1}^{n}log\left\{ \hat{f}(x_{i})\right\} .
\end{equation}
 Ivanov and Rozhkova (1981) have also used the leave-one-out approach
to obtain another KDE-based estimator of entropy similar to (11) with
its properties studied by Hall and Morton (1993).

Besides these Shannon entropy estimators so far discussed, where the
type of kernel function as well as the bandwidth size is not emphasized,
there are other estimators that have been introduced in the literature
based on KDE. They are presented in what follows.

Suppose $x_{1},\thinspace x_{2},\thinspace.\thinspace.\thinspace.,\thinspace x_{n}$
is a random sample of size $n$ from a continuous distribution $F(x)$
with probability density function $f(x)$. Alizadeh Noughabi (2010)
stated that for the entropy of the continuous random variable $X$,
$H(X)\thinspace=\thinspace-\int_{-\infty}^{\infty}f(x)log\left\{ f(x)\right\} dx\thinspace=\thinspace\int_{0}^{1}log\left\{ \frac{d}{dp}F^{-1}(p)\right\} dp$,
\begin{equation}
\frac{d}{dp}F^{-1}(p)\thinspace\simeq\thinspace\frac{X_{(i+m)}-X_{(i-m)}}{F\left(X_{(i+m)}\right)-F\left(X_{(i-m)}\right)}
\end{equation}
 where $\frac{i-1}{n}\thinspace\leq\thinspace p\thinspace\leq\thinspace\frac{i}{n};\thinspace i\thinspace=\thinspace m+1,\thinspace m+2,\thinspace.\thinspace.\thinspace.,\thinspace n-m$.
The work obtained an approximate equality of the denominator of (12)
and based on that, the author proposed a spacing--type kernel density
based estimator of entropy given by:
\[
HAN_{m,n}\thinspace=\thinspace-\frac{1}{n}\sum_{i=1}^{n}log\left\{ \frac{\hat{f}\left(X_{(i+m)}\right)-\hat{f}\left(X_{(i-m)}\right)}{2}\right\} 
\]
where $\hat{f}\left(x\right)$ is as defined in (8) and the kernel
function is chosen to be the standard normal density function with
a bandwidth $h\thinspace=\thinspace1.06sn^{-1/5};\thinspace s$ is
the sample standard deviation. He further proved that as $n,\thinspace m\thinspace\rightarrow\thinspace\infty,\thinspace m/n\thinspace\rightarrow\thinspace0$,
$HAN_{m,n}\thinspace\stackrel{Pr}{\rightarrow}\thinspace H(X)$ and
that the scale of the random variable $X$ has no effect on the accuracy
of the estimator.

In a yet another approximation of the denominator of (12), Zamanzade
and Arghami (2012) obtained 
\begin{equation}
F\left(X_{(i+m)}\right)-F\left(X_{(i-m)}\right)\thinspace=\thinspace\sum_{j=k_{1}(i)}^{k_{2}(i)-1}\left(\frac{f\left(X_{(j+1)}\right)+f\left(X_{(j)}\right)}{2}\right)\left(X_{(j+1)}-X_{(j)}\right)
\end{equation}
where $k_{1}(i)\thinspace=\thinspace\begin{cases}
\begin{array}{c}
1,\\
i-m,
\end{array} & \begin{array}{c}
if\thinspace i\leq m\\
if\thinspace i>m
\end{array}\end{cases}$ and $k_{2}(i)\thinspace=\thinspace\begin{cases}
\begin{array}{c}
i+m,\\
i-m,
\end{array} & \begin{array}{c}
if\thinspace i\leq n-m\\
if\thinspace i>n-m
\end{array}\end{cases}$. Estimating $f\left(X_{(j)}\right)$ by a kernel density estimation
method, they obtained two estimators of the Shannon entropy. They
are given by:
\[
HZA_{m,n}^{(1)}\thinspace=\thinspace\frac{1}{n}\sum_{i=1}^{n}log\left\{ b_{i}\right\} 
\]
and 
\[
HZA_{m,n}^{(2)}\thinspace=\thinspace\sum_{i=1}^{n}w_{i}log\left\{ b_{i}\right\} 
\]
where $b_{i}\thinspace=\thinspace\frac{X_{(i+m)}-X_{(i-m)}}{\sum_{j=k_{1}(i)}^{k_{2}(i)-1}\left(\frac{\hat{f}\left(X_{(j+1)}\right)+\hat{f}\left(X_{(j)}\right)}{2}\right)\left(X_{(j+1)}-X_{(j)}\right)}$;
$\hat{f}\left(X_{i}\right)\thinspace=\thinspace\frac{1}{nh}\sum_{j=1}^{n}k\left(\frac{X_{i}-X_{j}}{h}\right)$ 

and $w_{i}\thinspace=\thinspace\begin{cases}
\begin{array}{c}
\frac{m+i-1}{\sum_{i=1}^{n}w_{i}},\\
\frac{2m}{\sum_{i=1}^{n}w_{i}},\\
\frac{n-i+m}{\sum_{i=1}^{n}w_{i}},
\end{array} & \begin{array}{c}
1\leq i\leq m\\
m+1\leq i\leq n-m\\
n-m+1\leq i\leq n
\end{array}\end{cases}.$ They went further to prove that the accuracy of the estimator is
not affected by changes in the scale of the random variable. They
however did not show the consistency (or convergence) of the estimators.

Alizadeh Noughabi and Alizadeh Noughabi (2013) modified the slope
in Ebrahimi et al. (1994) as:
\begin{equation}
s_{i}(m,n)\thinspace=\thinspace\begin{cases}
\begin{array}{c}
\hat{f}\left(X_{(i)}\right),\\
\frac{2m/n}{X_{(i+m)}-X_{(i-m)}},\\
\hat{f}\left(X_{(i)}\right),
\end{array} & \begin{array}{c}
1\leq i\leq m\\
m+1\leq i\leq n-m\\
n-m+1\leq i\leq n
\end{array}\end{cases}.
\end{equation}
 where $\hat{f}\left(X_{(i)}\right)$ has the usual meaning. Based
on the modified slope, they proposed a new estimator of entropy of
a continuous random variable $X$ by:
\[
HAN_{m,n}^{2}\thinspace=\thinspace-\frac{1}{n}\sum_{i=1}^{n}log\left\{ s_{i}(n,m)\right\} 
\]
where $\hat{f}\left(X_{(i)}\right)$ in (14) is chosen to have the
standard normal density function as its kernel function with the bandwidth
$h$ chosen to be the normal optimal smoothing formula, $h\thinspace=\thinspace1.06sn^{-1/5}$;
$s$ is the sample standard deviation. Like most of the other estimators
based on spacing, they proved that the accuracy of the estimator is
not affected by changes in the scale of the random variable and that
as $n,\thinspace m\thinspace\rightarrow\thinspace\infty$, $\thinspace m/n\thinspace\rightarrow\thinspace0$,
$HAN_{m,n}^{2}\thinspace\stackrel{Pr}{\rightarrow}\thinspace H(X)$.

Bouzebda et al. (2013) expressed $\frac{d}{dp}F^{-1}(p)$ in (4) as
a quantile density function (qdf), $q(.)$. Using the kernel quantile
estimator (KQE) of Chen (1995), Falk (1986), and Cheng and Parzen
(1997), with some correction for boundary effect, they obtained a
new estimator of the Shannon entropy of a univariate random variable
$X$ as 
\[
HB_{\varepsilon,n}\thinspace=\thinspace\varepsilon log(\hat{q}(\varepsilon))\thinspace+\thinspace\varepsilon log(\hat{q}(1-\varepsilon))\thinspace+\thinspace\int_{\varepsilon}^{1-\varepsilon}log\left\{ \hat{q}_{n}(x)\right\} dx,\thinspace\varepsilon\in(0,\frac{1}{2})
\]
 where $\hat{q}_{n}(t)\thinspace=\thinspace\overline{q}(t)\thinspace+\thinspace$$\frac{1}{h}\left[K\left(\frac{t-1}{h}\right)X_{(n)}-K\left(\frac{t}{h}\right)X_{(1)}\right],\thinspace t\thinspace\in(0,1)$
and $\overline{q}(t)\thinspace=\thinspace\frac{1}{h}\sum_{i=1}^{n-1}K\left(\frac{t-i/n}{h}\right)\left(X_{(i+1)}-X_{(i)}\right),\thinspace t\thinspace\in(0,1)$.
They went further to compare its mean square error (MSE) and bias
with those of some other estimators and concluded that the estimator
was efficient.

Bouzebda and Elhattab (2014) tried to simplify the Bouzebda et al.
(2013) by obtaining an expression for the part resulting in numerical
integration. As a result, they obtained
\[
HBE_{\varepsilon,n}\thinspace=\thinspace\varepsilon log(\hat{q}(\varepsilon))\thinspace+\thinspace\varepsilon log(\hat{q}(1-\varepsilon))\thinspace+\thinspace\frac{1}{n}\sum_{i=\left[\varepsilon n\right]}^{\left[(1-\varepsilon)n\right]}log\left\{ \hat{q}(F_{n}(x_{i}))\right\} ,\thinspace\varepsilon\in(0,1)
\]
 where $\hat{q}(t)\thinspace=\thinspace\overline{q}(t)\thinspace+\thinspace$$\frac{1}{h}\left[K\left(\frac{t-1}{h}\right)X_{(n)}-K\left(\frac{t}{h}\right)X_{(1)}\right],\thinspace t\thinspace\in(0,1)$;
$\overline{q}(t)\thinspace=\thinspace\frac{1}{h}\sum_{i=1}^{n-1}K\left(\frac{t-i/n}{h}\right)\left(X_{(i+1)}-X_{(i)}\right),\thinspace t\thinspace\in(0,1)$;
and $F_{n}(t)\thinspace=\thinspace\frac{1}{n}\sum_{i=1}^{n}1_{(-\infty,t]}(X_{i}),\thinspace t\thinspace\in R$;
$1_{A}$ is the indicator function of the event $A$. They went further
to state that the new estimator reduces the computational requirement
of $HB_{\varepsilon,n}$ that can be very significant when $n$ is
large. It is important to note here that both $HB_{\varepsilon,n}$
and $HBE_{n,\varepsilon}$ are based on the KDE of quantile density
function, qdf, instead of probability density function, pdf, as seen
in others in this category.

\subsection{Estimators based on $\boldsymbol{k}-$nearest neighbour estimation
method}

Nearest neighbour concept is generally defined for a $d-$dimensional
space, $d\thinspace\geq\thinspace1$, and so it is presented in this
work. Suppose a continuous random vector $\boldsymbol{X}$ has an
unknown distribution function with a pdf, $f(\boldsymbol{X})$. Let
$\boldsymbol{X}_{1},\thinspace\boldsymbol{X}_{2},\thinspace.\thinspace.\thinspace.,\thinspace\boldsymbol{X}_{n}$
be a random sample of size $n$ observation vectors from the distribution
and let $\varDelta_{k}^{d}(\boldsymbol{X}_{i})$ be the nearest neighbour
distance between an observation vector $\boldsymbol{X}_{i}$ and its
$k$th nearest neighbour, ($kNN$), among the sample. Then,
\begin{equation}
\varDelta_{k}^{d}(\boldsymbol{X}_{i})\thinspace=\thinspace\underset{j\neq i,\,j\leq n}{min_{k}}\left\Vert \boldsymbol{X}_{i}-\boldsymbol{X}_{j}\right\Vert 
\end{equation}
where $\left\Vert .\right\Vert $ is any distance measure, usually
the Euclidean or Manhattan distance. This metric, no doubt, is a ball
which is centred at $\boldsymbol{X}_{i}$ with radius, $r\thinspace=\thinspace\varDelta_{k}^{d}(\boldsymbol{X}_{i})$,
i.e $B\left(\boldsymbol{X}_{i},\varDelta_{k}^{d}(\boldsymbol{X}_{i})\right)$
and volume, $V_{k}(\boldsymbol{X}_{i})$.

The $k-$nearest neighbour distance in (15) has been used traditionally
to estimate the pdf, $f(\boldsymbol{X})$ of the random vector $\boldsymbol{X}$
from the sample, see for instance, Loftsgaarden and Quesenberry (1965).
Later, it became a powerful instrument for the estimation of the Shannon
entropy of random variables. Penrose and Yukich (2013) proved the
limit behaviour of the estimators ($H_{n}(\boldsymbol{X})$) in the class under $m-$dimensional
manifold and showed that they are such that 
\begin{equation}
n^{-1/2}\left[H_{n}(\boldsymbol{X}-E\left(H_{n}(\boldsymbol{X}\right)\right]\thinspace\stackrel{D}{\rightarrow}\thinspace N(0,\psi)
\end{equation}
where $\psi$ is the limit variance of $H_{n}(\boldsymbol{X}$ as $n\thinspace\rightarrow\thinspace\infty$.

The first known estimator of the Shannon differential entropy via
the $kNN$ approach (Charzynska \& Gambin, 2015) is due to Kozachenko
and Leonenko (1987) which they gave for a $d-$dimensional space,
$d\thinspace\geq\thinspace1$. For a random sample of $n$ independent
observation vectors from a continuous distribution with pdf, $f(\boldsymbol{X})$,
let $\varDelta^{d}(\boldsymbol{X}_{i})$ be the distance function
of the observation $\boldsymbol{X}_{i}$ to its nearest neighbour
$\boldsymbol{X}_{j}$ in the sample and let $V(\boldsymbol{X}_{i})$
be the volume of a ball, $B\left(\boldsymbol{X}_{i},\varDelta^{d}(\boldsymbol{X}_{i})\right)$
defined for the point$\boldsymbol{X}_{i}$. Kozachenko and Leonenko
(1987) obtained the estimator of entropy as 
\[
HKL_{n}\thinspace=\thinspace-\frac{1}{n}\sum_{i=1}^{n}log\left\{ f_{n}(\boldsymbol{X}_{i})\right\} \thinspace+\thinspace\gamma
\]
 where $f_{n}(\boldsymbol{X}_{i})$ is the $kNN$ estimator of the
$f(\boldsymbol{X})$ for $k\thinspace=\thinspace1$, given by $f_{n}(\boldsymbol{X}_{i})\thinspace=\thinspace\left[(n-1)\varDelta^{d}(\boldsymbol{X}_{i})V(\boldsymbol{X}_{i})\right]^{-1}$
and $\gamma$ is the Euler constant which is approximately 0.5772.
Kozachenko and Leonenko (1987) showed that the estimator
converges in probability to the true population entropy for large
$n$, i.e. $HKL_{n}\thinspace\stackrel{P}{\rightarrow}\thinspace H(X)$
as $n\thinspace\rightarrow\thinspace\infty$.

Shannon (1948) had stated that if the logarithmic base in (1) and
(2) is 2, the entropy measure will represent the amount of bits of
generated information per symbol or per second. However, Keshmiri
(2020) also stated that the choice of logarithmic base does not necessarily
impose any restrictions to the development of entropy as long as the
computed entropy can be conveniently interpreted in the context of
its application. Developing an entropy estimator from the standpoint
of bits per second of generated information, Victor (2002) obtained
another $kNN$ estimator for $k\thinspace=\thinspace1$, which is
simmilar to the Kozachenko and Leonenko (1987). It is given by
\[
HV_{n}\thinspace=\thinspace\frac{d}{n}\sum_{i=1}^{n}log_{2}\left\{ \varDelta^{d}(\boldsymbol{X}_{i})\right\} \thinspace+\thinspace log_{2}\left\{ V(\boldsymbol{X}_{i})(n-1)\right\} \thinspace+\thinspace\frac{\gamma}{ln(2)}
\]
where $log_{2}\left\{ .\right\} $ is the logarithmic function to
base 2, $V(\boldsymbol{X}_{i})\thinspace=\thinspace\frac{d\pi^{d/2}}{\varGamma\left(\frac{d}{2}+1\right)}$
and all the other terms have their usual meanings.

With $\left\{ \boldsymbol{X}_{i}\right\} _{i=1,2,...,n}$ as a random
sample of $n$ independent observations from an unknown distribution
having pdf, $f(\boldsymbol{X})$, which has $\varDelta_{k}^{d}(X_{i})$
as the usual Euclidean distance between an observation $\boldsymbol{X}_{i}$
and its $k$th nearest neighbour in the sample whose volume of a
shere with radius, $\varDelta_{k}^{d}(\boldsymbol{X}_{i})$ is $\left[\pi^{d/2}\varDelta_{k}^{d}(\boldsymbol{X}_{i})\right]/\varGamma\left(\frac{d}{2}+1\right)$,
Singh et al. (2003) obtained a relation for the $kNN$ density estimator,
$f_{n}(\boldsymbol{X})$ as
\begin{equation}
f_{n}(\boldsymbol{X}_{i})\left(\frac{\pi^{d/2}\varDelta_{k}^{d}(\boldsymbol{X}_{i})}{\varGamma\left(\frac{d}{2}+1\right)}\right)\thinspace=\thinspace\frac{k}{n}
\end{equation}
Using the relation in (17), they obtained an asymptotically unbiased
estimator of the Shannon differential entropy, similar to the Kozachenko
and Leonenko (1987) estimator as
\[
HS_{k,n}\thinspace=\thinspace\frac{d}{n}\sum_{i=1}^{n}log\left\{ \varDelta_{k}^{d}(\boldsymbol{X}_{i})\right\} \thinspace+\thinspace log(n)\thinspace+\thinspace log\left\{ \frac{\pi^{d/2}}{\varGamma\left(\frac{d}{2}+1\right)}\right\} \thinspace+\thinspace\gamma\thinspace-\thinspace L_{k-1}
\]
where $L_{k}\thinspace=\thinspace\sum_{i=1}^{n}\frac{1}{i};\thinspace L_{0}\thinspace=\thinspace0$.
They proved the consistency of the estimator and initiated the comparison
of the asymptotic variance of the $kNN$ entropy estimators and those
of the $m-$spacing.

Nilsson and Kleijn (2004) obtained an estimator of the Shannon differential
entropy in a $\tilde{d}-$dimentional space, $\tilde{d}\in\mathbb{Z}$.
They used the process of high rate quantization that accounts for
embedded manifolds in $\mathbb{R}^{d}$ ($\tilde{d}\thinspace\leq\thinspace d)$,
to obtain a $kNN$ estimator of the entropy that is similar to the
Kozachenko and Leonenko (1987). It is obtained from the standpoint
of entropy as a measure of bits per second and as a result, defined
for $d$--dimensional space as
\[
HN_{n}\thinspace=\thinspace\frac{d}{n}\sum_{i=1}^{n}log_{2}\left\{ \varDelta^{d}(\boldsymbol{X}_{i})\right\} \thinspace+\thinspace log_{2}(n)\thinspace+\thinspace\frac{\gamma}{ln(2)}\thinspace+\thinspace\frac{1}{2}log_{2}(2e)
\]
where $\gamma$ remains the Euler constant, approximated by 0.5772
and $\varDelta^{d}(\boldsymbol{X}_{i})$ is the $k$th nearest neighbour
distance, $k\thinspace=\thinspace1$, for $\boldsymbol{X}_{i}$. The
$HN_{n}$ estimator was developed to allow fractional manifold dimensions,
$\tilde{d}\in\mathbb{Z}$.

Kraskov et al. (2004) studied the $HKL_{n}$ estimator of Kozachenko
and Leonenko (1987) where the $kNN$ estimator of the $f(\boldsymbol{X})$
is for $k\thinspace=\thinspace1$. They extended the same estimator
for any $k\thinspace\geq1$ to obtain
\[
HK_{n}\thinspace=\thinspace-\varPsi(k)\thinspace+\thinspace\varPsi(n)\thinspace+\thinspace log\left\{ V(\boldsymbol{X}_{i})\right\} \thinspace+\thinspace\frac{d}{n}\sum_{i=1}^{n}log\left\{ 2\varDelta_{k}^{d}(\boldsymbol{X}_{i})\right\} 
\]
where $\varPsi(x)\thinspace=\thinspace\nicefrac{\varGamma^{\prime}(x)}{\varGamma(x)}$
is the digamma function, $V(\boldsymbol{X}_{i})$ is the unit volume
of a ball, $B\left(\boldsymbol{X}_{i},\varDelta_{k}^{d}(\boldsymbol{X}_{i})\right)$
defined for the point $\boldsymbol{X}_{i}$, which they gave as $\nicefrac{2^{d}\pi^{d/2}}{\varGamma(1+\frac{d}{2})}$
for the Euclidean distance and $\varDelta_{k}^{d}(x_{i})$ remains
the distance function of the observation $\boldsymbol{X}_{i}$ to
its $k$th nearest neighbour. They stated that the estimator is asymptotically
unbiased.

Leonenko et al. (2008) obtained a $kNN$ estimator of the Renyi $H_{\alpha}^{*}(X)$
and Tsallis $H_{\alpha}(X)$ entropies (Renyi, 1961; Tsallis, 1988).
It is well known that the Renyi entropy
\begin{equation}
H_{\alpha}^{*}(X)\thinspace=\thinspace\frac{1}{1-\alpha}log\int_{-\infty}^{\infty}f^{\alpha}(x)dx;\thinspace\alpha\thinspace\neq\thinspace1
\end{equation}
and the Tsallis entropy
\begin{equation}
H_{\alpha}(X)\thinspace=\thinspace\frac{1}{\alpha-1}\left(1-\int_{-\infty}^{\infty}f^{\alpha}(x)dx\right);\thinspace\alpha\thinspace\neq\thinspace1
\end{equation}
 of a random variable $X$ which has distribution function with probability
density function $f(x)$ both tend to the Shannon entropy $H(X)$
as $\alpha$ tends to 1. As a result, they took the limit of their
estimator as $\alpha\thinspace\rightarrow\thinspace1$ to obtain an
estimator of the Shannon entropy. This coincides with the estimator due to Goria et al. (2005), and it is given by
\[
HL_{n}\thinspace=\thinspace\frac{d}{n}\sum_{i=1}^{n}log\left\{ \xi_{n,i,k}\right\} 
\]
where $\xi_{n,i,k}\thinspace=\thinspace(n-1)exp\left[-\varPsi(k)\right]V(\boldsymbol{X}_{i})\varDelta_{k}^{d}(\boldsymbol{X}_{i})$
for a random vector $\boldsymbol{X}$. In a more computational and
less compact form, the $HL_{n}$ estimator can be presented as 
\[
HL_{n}\thinspace=\thinspace-\varPsi(k)\thinspace+\thinspace log(n-1)\thinspace+\thinspace log\left\{ V(\boldsymbol{X}_{i})\right\} \thinspace+\thinspace\frac{d}{n}\sum_{i=1}^{n}log\left\{ \varDelta_{k}^{d}(\boldsymbol{X}_{i})\right\} .
\]

In addition to the estimators described above, other variants have
been presented in the literature. For instance, Charzynska and Gambin
(2015) obtained a bias correction factor for the estimators under
high dimensions. Lombardi and Pant (2016) argued that the pdf, $f(\boldsymbol{X})$,
is not uniform inside the closed ball, $B(\boldsymbol{X_{i}},\varDelta_{k}^{d}(\boldsymbol{X}_{i}))$,
as assumed by Kozachenko and Leonenko (1987). As a result, they obtained
a correction factor for the non-uniformity of the pdf inside the ball.
Lord et al. (2018) defined entropy from the point of geometry. They
obtained a function of the Euclidean distance between a point $\boldsymbol{X}_{i}$
and its $k$th nearest neighbour to be lengths of axes of the ellipsoid
centred at the $\boldsymbol{X}_{i}$ with a volume $V(\boldsymbol{X}_{i})$.
From this standpoint, they obtained a variant of the Kozachenko and
Leonenko (1987) estimator. Weighted estimators of the class has also
been considered by Sricharan et al. (2013), Moon (2016) as well as
Berrett et al. (2019). The estimators in the class has equally been
developed for some special class of distributions. Such estimators
include Misra et al. (2010).

Some important asymptotic properties of the setimators in this class
have been extensively studied. They include their rate of convergence,
asymptotic variance, mean square, unbiasedness, asymptotic distribution
and consistency. Kaltchenko and Timofeeva (2010) studied their rate
of convergence and showed that they converge at a rate slower than
$O\left(1/logn\right)$. Biau and Devroye (2015) showed that the bias
of the class of estimators vanishes asymptotically at low dimensions
($d\leq3)$. Bulsinski and Dimitrov (2018) proved their asymptotic
unbiasedness while Devroye and Gyorfi (2022) proved their consistency.

Besides the three entropy estimation techniques so far discussed in
this work, some few other methods, although less popular, have been
used to estimate the differential entropy. For instance, Menez et
al. (2008) have used the principle of minimum description length (MDL)
according to Rissanen (2007) to propose a histogram-type estimator.
Kohansal and Rezakhah (2015) used moving average (MA) technique to
transform the observations in one--dimensional case before applying
the spacing method to obtain two estimators of Shannon differential
entropy of univariate absolutely continuous random variables. Also,
Hino et al. (2015) used simple linear regression to estimate the values
of density function and its second derivatives at a point in order
to obtain four different estimators of the Shannon differential entropy
based on second order expansion of the probability mass of a ball
around a point.

\section{Simulation Studies}

In this section, the estimators of the Shannon differential entropy
are compared through Monte Carlo simulations. The estimators in each
of the three classes are first compared within themselves using two
perfomance measures of empirical root mean square error (RMSE) and
empirical bias. Suppose $H_{n}(X)$ is an estimator of the Shannon
differential entropy $H(X)$. The empirical RMSE of $H_{n}(X)$ is
defined by: 
\begin{equation}
ERMSE\left(H_{n}(X)\right)\thinspace=\thinspace\sqrt{\frac{1}{N}\sum_{j=1}^{N}\left(H_{n}(X)\thinspace-\thinspace H(X)\right)^{2}}
\end{equation}
 Also, the absolute empirical bias is give by:
\begin{equation}
abs\left(Ebias\left(H_{n}(X)\right)\right)\thinspace=\thinspace\left|\frac{1}{N}\sum_{j=1}^{N}H_{n}(X)\thinspace-\thinspace H(X)\right|
\end{equation}
 where $N$ is the number of samples of size $n$ each from which
$H_{n}(X)$ is obtained. Using these performance indices, four estimators
are selected from the 16 considered estimators based on window size
spacings while three each are selected from the five and six estimators
considered from the classes based on kernel density estimation (KDE)
and $k-$nearest neighbour ($k$NN) method respectively.

Throughout the simulation studies, three $d-$dimensional statistical
distributions are considered, $d\thinspace=\thinspace1,\thinspace2,\thinspace3,\thinspace.\thinspace.\thinspace.$.
They include the $d-$dimensional normal, with support $R^{d}$; the
$d-$dimensional exponential, with support $[0,\infty)^{d}$ and the
$d-$dimensional uniform, with support $[0,1]^{d}$. These distributions
are selected to cover the three main supports for distributional laws,
in one part, and also to cover symmetric and skewed distributions
on the other part. Precisely, 10,000 samples of sizes $n\thinspace=\thinspace10,\thinspace50,\thinspace100$
each are independently simulated from each of the $d-$dimensional
standard normal, $d-$dimensional standard exponential and $d-$dimensional
uniform distributions respectively for $d\thinspace=\thinspace1,\thinspace2,\thinspace3,\thinspace5$
and the entropy estimates from each of all the considered estimators
are computed, giving rise to a total of 10,000 computed estimates
of each estimator for each combination of sample size, $n$ and variable
dimension, $d$ under each distribution. Then in each case, the empirical
RMSE and absolute bias of each estimator is obtained using (20) and
(21) respectively. The results are presented in Tables 1 - 3 for the
spacings based estimators, respectively for each of the distributions
considered, Tables 4 - 6 for the KDE based estimators, respectively
for each of the sample sizes $n\thinspace=\thinspace10,\thinspace50,\thinspace100$,
and Tables 7 - 9 for the $k$NN based estimators, respectively for
each of the three distributions considered.

The spacings based estimators are tractable only for one-dimensional
observations. As a result, the 16 estimators considered in this class
are done only at $d\thinspace=\thinspace1.$ Also, it is known that
the performance of the estimators in this class is affected by the
window size, $(m)$, $m\thinspace\leq\thinspace\frac{n}{2}$. It has
been shown that the optimal $m$ which is given by $m^{*}\thinspace=\thinspace\left[\sqrt{n}\thinspace+\thinspace0.5\right]$
according to Grzegorzewski and Wieczorkowski (1999) does not always
hold for all the estimators across all distributional laws, see for
instance, Al-Omari (2016). A more accommodating optimal $m$ is introduced
in this work. It is given by:
\begin{equation}
m^{+}\thinspace=\thinspace m^{*}\thinspace\pm2
\end{equation}
 This means that for a sample size of $n\thinspace=\thinspace10$,
for instance, with $m^{*}\thinspace=\thinspace\left[\sqrt{10}\thinspace+\thinspace0.5\right]\thinspace=\thinspace3$,
$m^{+}\thinspace=\thinspace3\thinspace\pm\thinspace2\thinspace=\thinspace[1,\thinspace5]$.
This newly introduced window size interval is used throughout in this
study and the best is obtained in each case as the one with smallest
RMSE among the ones with small bias values, for instance.

Again, the class of estimators based on KDE is known to depend on
the type of kernel used as well as the bandwidth, $h$ (for $d\thinspace=\thinspace1$)
and $H$ (for $d\thinspace>\thinspace1$). Throughout the computations
in this class, the Gaussian kernel is used with a normal scale bandwidth
of $h\thinspace=\thinspace1.06n\hat{\sigma}^{-1/5}$ (for $h\thinspace=\thinspace1$)
and $H\thinspace=\thinspace\left[\frac{4}{n(d+2)}\right]^{\frac{2}{d+4}}\hat{\sum}$
(for $d\thinspace>\thinspace1$) which are known to be high performing
candidates according to Wand and Jones (1995), Hardle and Muller (2004),
Chacon and Duong (2015) and Gramacki (2018), where $\hat{\sigma}$
and $\hat{\sum}$ are the estimators of the variance of the random
variable, $X$ and the covariance matrix of the random vector, $\boldsymbol{X}$,
respectively, whose entropy is being estimated. The computations were
conducted with the help of the $R-$statistical software package,
$ks$. Except for Ahmad and Lin (1975) estimator, $HAL_{n}$, all
the estimators of the Shannon entropy considered in this class are
compared at $d\thinspace=\thinspace1$ only. This is because the window
size ($m$) is introduced in all of them except the former.

In the $k-$nearest neighbour class of entropy estimators, it is known
that their performance is affected by the distance function, $\varDelta_{k}^{d}(\boldsymbol{X}_{i})$
used as well as the number of nearest neighbours, $k$ ($k\thinspace\leq\thinspace n-1$).
In line with Singh et al. (2003), all the Shannon entropy estimators
considered in this class are estimated for $k\thinspace=\thinspace1,\thinspace3,\thinspace5$
respectively with the Euclidean distance as the distance function
and computations were done with the help of the $R-$statistical software
package, dbscan.

\section{Discussion of Results}

From Tables 1 - 3, it is shown that the optimal $m$ for a fixed $n$
that minimized the RMSE and absolute bias of the estimators within
the optimal interval, $m^{+}$ is not the same across the estimators.
Also, the optimal $m$ within the $m^{+}$ for a fixed $n$ is dependent
on the performance measure of interest. In other words, the $m$ that
minimized RMSE need not be the same that minimized the bias. As expected,
both the RMSE and bias decreased steadily with increasing sample size
in all the estimators considered in the spacings based estimators.
Again, they all maintained relatively low bias values across the sample
sizes and distributions with somewhat differences in RMSE values across
distributions. This is because estimators with least RMSE values under
a distribution are observed to differ somewhat from those of other
distribution(s) considered. Again, it is observed that different estimators
showed least RMSE values at different sample sizes under the same
distribution. As a result, the best five estimators in each case of
sample size and distribution in terms of RMSE values with reduced
bias are presented in Table 10.
\begin{table}
\caption{RMSE and absolute bias of the spacing based Shannon entropy estimators
under the uniform distribution, $U(0,1)$, $H(X)\thinspace=\thinspace0$}
\small
\begin{centering}
\begin{tabular}{cccccccccccccc}
\hline 
\multirow{2}{*}{$H_{n}(X)$} & \multicolumn{2}{c}{$n=10\thinspace(3)$} & \multicolumn{2}{c}{$n=50\thinspace(7)$} & \multicolumn{2}{c}{$n=100\thinspace(10)$} & \multirow{2}{*}{$H_{n}(X)$} & \multicolumn{2}{c}{$n=10\thinspace(3)$} & \multicolumn{2}{c}{$n=50\thinspace(7)$} & \multicolumn{2}{c}{$n=100\thinspace(10)$}\tabularnewline
\cline{2-7} \cline{3-7} \cline{4-7} \cline{5-7} \cline{6-7} \cline{7-7} \cline{9-14} \cline{10-14} \cline{11-14} \cline{12-14} \cline{13-14} \cline{14-14} 
 & $R$ & $B$ & $R$ & $B$ & $R$ & $B$ &  & $R$ & $B$ & $R$ & $B$ & $R$ & $B$\tabularnewline
\hline 
$HV_{m,n}$ & 0.571 & 0.399 & 0.147 & 0.126 & 0.097 & 0.091 & $HZ_{m,n}$ & 0.447 & 0.157 & 0.133 & 0.187 & 0.138 & 0.167\tabularnewline
$HV_{m,n}$ & 0.453 & 0.226 & 0.151 & 0.170 & 0.100 & 0.103 & $HZ_{m,n}$ & 0.180 & 0.002 & 0.192 & 0.242 & 0.168 & 0.183\tabularnewline
$HV_{m,n}$ & 0.453 & 0.471 & 0.155 & 0.222 & 0.103 & 0.127 & $HZ_{m,n}$ & 0.339 & 0.452 & 0.250 & 0.254 & 0.197 & 0.215\tabularnewline
$HV_{m,n}$ & 0.483 & 0.387 & 0.163 & 0.137 & 0.107 & 0.148 & $HZ_{m,n}$ & 0.592 & 0.604 & 0.307 & 0.287 & 0.226 & 0.236\tabularnewline
$HV_{m,n}$ & 0.532 & 0.834 & 0.171 & 0.126 & 0.111 & 0.114 & $HZ_{m,n}$ & 0.857 & 1.057 & 0.363 & 0.365 & 0.255 & 0.214\tabularnewline
$HE_{m,n}$ & 0.296 & 0.026 & 0.058 & 0.124 & 0.033 & 0.030 & $HA_{m,n}^{(1)}$ & 0.518 & 0.571 & 0.093 & 0.041 & 0.054 & 0.024\tabularnewline
$HE_{m,n}$ & 0.220 & 0.553 & 0.057 & 0.019 & 0.033 & 0.011 & $HA_{m,n}^{(1)}$ & 0.350 & 0.493 & 0.085 & 0.135 & 0.050 & 0.083\tabularnewline
$HE_{m,n}$ & 0.214 & 0.056 & 0.057 & 0.013 & 0.033 & 0.000 & $HA_{m,n}^{(1)}$ & 0.295 & 0.309 & 0.079 & 0.038 & 0.048 & 0.028\tabularnewline
$HE_{m,n}$ & 0.225 & 0.075 & 0.060 & 0.050 & 0.034 & 0.040 & $HA_{m,n}^{(1)}$ & 0.279 & 0.323 & 0.075 & 0.001 & 0.046 & 0.036\tabularnewline
$HE_{m,n}$ & 0.227 & 0.402 & 0.062 & 0.079 & 0.035 & 0.024 & $HA_{m,n}^{(1)}$ & 0.270 & 0.003 & 0.073 & 0.022 & 0.044 & 0.060\tabularnewline
$HE_{m,n}^{(1)}$ & 0.453 & 0.463 & 0.077 & 0.051 & 0.044 & 0.032 & $HA_{m,n}^{(2)}$ & 0.486 & 0.497 & 0.062 & 0.032 & 0.029 & 0.036\tabularnewline
$HE_{m,n}^{(1)}$ & 0.280 & 0.030 & 0.069 & 0.021 & 0.041 & 0.020 & $HA_{m,n}^{(2)}$ & 0.290 & 0.103 & 0.050 & 0.023 & 0.024 & 0.001\tabularnewline
$HE_{m,n}^{(1)}$ & 0.232 & 0.041 & 0.063 & 0.045 & 0.038 & 0.012 & $HA_{m,n}^{(2)}$ & 0.220 & 0.007 & 0.042 & 0.023 & 0.021 & 0.008\tabularnewline
$HE_{m,n}^{(1)}$ & 0.215 & 0.098 & 0.059 & 0.042 & 0.036 & 0.033 & $HA_{m,n}^{(2)}$ & 0.183 & 0.363 & 0.037 & 0.022 & 0.020 & 0.029\tabularnewline
$HE_{m,n}^{(1)}$ & 0.207 & 0.018 & 0.056 & 0.042 & 0.034 & 0.011 & $HA_{m,n}^{(2)}$ & 0.166 & 0.139 & 0.036 & 0.106 & 0.020 & 0.012\tabularnewline
$HE_{m,n}^{(2)}$ & 0.266 & 0.103 & 0.066 & 0.056 & 0.040 & 0.023 & $HA_{m,n}^{(3)}$ & 0.451 & 0.390 & 0.065 & 0.043 & 0.034 & 0.066\tabularnewline
$HE_{m,n}^{(2)}$ & 0.170 & 0.039 & 0.059 & 0.055 & 0.037 & 0.012 & $HA_{m,n}^{(3)}$ & 0.271 & 0.577 & 0.055 & 0.008 & 0.030 & 0.077\tabularnewline
$HE_{m,n}^{(2)}$ & 0.141 & 0.079 & 0.054 & 0.035 & 0.035 & 0.032 & $HA_{m,n}^{(3)}$ & 0.209 & 0.078 & 0.047 & 0.029 & 0.026 & 0.004\tabularnewline
$HE_{m,n}^{(2)}$ & 0.138 & 0.036 & 0.050 & 0.022 & 0.033 & 0.035 & $HA_{m,n}^{(3)}$ & 0.185 & 0.076 & 0.043 & 0.023 & 0.024 & 0.011\tabularnewline
$HE_{m,n}^{(2)}$ & - & - & 0.047 & 0.075 & 0.031 & 0.033 & $HA_{m,n}^{(3)}$ & 0.171 & 0.045 & 0.039 & 0.026 & 0.022 & 0.041\tabularnewline
$HC_{m,n}$ & 0.449 & 0.310 & 0.067 & 0.010 & 0.038 & 0.003 & $HB_{m,n}^{(1)}$ & - & - & 0.041 & 0.013 & 0.024 & 0.010\tabularnewline
$HC_{m,n}$ & 0.291 & 0.552 & 0.072 & 0.023 & 0.043 & 0.011 & $HB_{m,n}^{(1)}$ & 0.270 & 0.182 & 0.040 & 0.038 & 0.031 & 0.002\tabularnewline
$HC_{m,n}$ & 0.301 & 0.205 & 0.074 & 0.049 & 0.047 & 0.089 & $HB_{m,n}^{(1)}$ & 0.175 & 0.042 & 0.051 & 0.042 & 0.040 & 0.041\tabularnewline
$HC_{m,n}$ & 0.309 & 0.392 & 0.085 & 0.085 & 0.050 & 0.079 & $HB_{m,n}^{(1)}$ & 0.172 & 0.180 & 0.067 & 0.039 & 0.049 & 0.075\tabularnewline
$HC_{m,n}$ & 0.339 & 0.274 & 0.087 & 0.030 & 0.053 & 0.088 & $HB_{m,n}^{(1)}$ & 0.228 & 0.001 & 0.084 & 0.095 & 0.058 & 0.056\tabularnewline
$HW_{m,n}^{(1)}$ & 0.237 & 0.341 & 0.288 & 0.218 & 0.327 & 0.350 & $HB_{m,n}^{(2)}$ & - & - & 0.041 & 0.014 & 0.025 & 0.018\tabularnewline
$HW_{m,n}^{(1)}$ & 0.198 & 0.115 & 0.402 & 0.426 & 0.395 & 0.414 & $HB_{m,n}^{(2)}$ & 0.232 & 0.024 & 0.041 & 0.029 & 0.032 & 0.014\tabularnewline
$HW_{m,n}^{(1)}$ & 0.466 & 0.329 & 0.524 & 0.433 & 0.466 & 0.463 & $HB_{m,n}^{(2)}$ & 0.145 & 0.020 & 0.053 & 0.050 & 0.040 & 0.007\tabularnewline
$HW_{m,n}^{(1)}$ & 0.912 & 0.825 & 0.654 & 0.670 & 0.538 & 0.533 & $HB_{m,n}^{(2)}$ & 0.167 & 0.022 & 0.069 & 0.102 & 0.049 & 0.035\tabularnewline
$HW_{m,n}^{(1)}$ & 1.428 & 1.352 & 0.791 & 0.787 & 0.613 & 0.637 & $HB_{m,n}^{(2)}$ & 0.227 & 0.201 & 0.087 & 0.076 & 0.058 & 0.077\tabularnewline
$HP_{m,n}$ & 0.355 & 0.116 & 0.062 & 0.084 & 0.032 & 0.033 & $HJ_{m,n}$ & 0.374 & 0.209 & 0.059 & 0.084 & 0.054 & 0.065\tabularnewline
$HP_{m,n}$ & 0.236 & 0.190 & 0.052 & 0.007 & 0.027 & 0.021 & $HJ_{m,n}$ & 0.202 & 0.289 & 0.072 & 0.105 & 0.060 & 0.068\tabularnewline
$HP_{m,n}$ & 0.200 & 0.176 & 0.045 & 0.055 & 0.024 & 0.003 & $HJ_{m,n}$ & 0.178 & 0.034 & 0.082 & 0.067 & 0.066 & 0.072\tabularnewline
$HP_{m,n}$ & 0.177 & 0.049 & 0.039  & 0.036 & 0.021 & 0.008 & $HJ_{m,n}$ & 0.200 & 0.148 & 0.098 & 0.089 & 0.068 & 0.069\tabularnewline
$HP_{m,n}$ & 0.172 & 0.030 & 0.036 & 0.060 & 0.020 & 0.009 & $HJ_{m,n}$ & 0.255 & 0.407 & 0.110 & 0.091 & 0.070 & 0.071\tabularnewline
$HA_{m,n}$ & 0.445 & 0.167 & 0.039 & 0.059 & 0.026 & 0.004 & $HM_{m,n}$ & 1.154 & 1.120 & 0.097 & 0.076 & 0.052 & 0.041\tabularnewline
$HA_{m,n}$ & 0.226 & 0.116 & 0.043 & 0.032 & 0.034 & 0.034 & $HM_{m,n}$ & - & - & 0.076 & 0.049 & 0.048 & 0.055\tabularnewline
$HA_{m,n}$ & 0.162 & 0.168 & 0.057 & 0.010 & 0.043 & 0.021 & $HM_{m,n}$ & 0.258 & 0.432 & 0.066 & 0.040 & 0.044 & 0.035\tabularnewline
$HA_{m,n}$ & 0.190 & 0.093 & 0.073 & 0.059 & 0.052 & 0.032 & $HM_{m,n}$ & 0.175 & 0.195 & 0.061 & 0.053 & 0.043 & 0.018\tabularnewline
$HA_{m,n}$ & 0.252 & 0.165 & 0.090 & 0.061 & 0.061 & 0.074 & $HM_{m,n}$ & 0.168 & 0.126 & 0.059 & 0.021 & 0.041 & 0.021\tabularnewline
\hline 
\multicolumn{14}{c}{$R$ = RMSE, $B$ = $\left|Bias\right|$}\tabularnewline
\end{tabular}
\par\end{centering}
\end{table}

\begin{table}
\caption{RMSE and absolute bias of the spacing based Shannon entropy estimators
under the standard normal distribution, $H(X)\thinspace=\thinspace1.4189$}
\small
\begin{centering}
\begin{tabular}{cccccccccccccc}
\hline 
\multirow{2}{*}{$H_{n}(X)$} & \multicolumn{2}{c}{$n=10\thinspace(3)$} & \multicolumn{2}{c}{$n=50\thinspace(7)$} & \multicolumn{2}{c}{$n=100\thinspace(10)$} & \multirow{2}{*}{$H_{n}(X)$} & \multicolumn{2}{c}{$n=10\thinspace(3)$} & \multicolumn{2}{c}{$n=50\thinspace(7)$} & \multicolumn{2}{c}{$n=100\thinspace(10)$}\tabularnewline
\cline{2-7} \cline{3-7} \cline{4-7} \cline{5-7} \cline{6-7} \cline{7-7} \cline{9-14} \cline{10-14} \cline{11-14} \cline{12-14} \cline{13-14} \cline{14-14} 
 & $R$ & $B$  & $R$ & $B$  & $R$ & $B$  &  & $R$ & $B$  & $R$ & $B$  & $R$ & $B$ \tabularnewline
\hline 
$HV_{m,n}$ & 0.673 & 0.672 & 0.195 & 0.017 & 0.120 & 0.181 & $HZ_{m,n}$ & 0.556 & 0.746 & 0.153 & 0.156 & 0.158 & 0.155\tabularnewline
$HV_{m,n}$ & 0.595 & 0.418 & 0.196 & 0.121 & 0.121 & 0.056 & $HZ_{m,n}$ & 0.299 & 0.332 & 0.204 & 0.145 & 0.187 & 0.214\tabularnewline
$HV_{m,n}$ & 0.619 & 1.095 & 0.198 & 0.075 & 0.121 & 0.014 & $HZ_{m,n}$ & 0.313 & 0.303 & 0.255 & 0.267 & 0.218 & 0.170\tabularnewline
$HV_{m,n}$ & 0.663 & 0.581 & 0.202 & 0.163 & 0.122 & 0.028 & $HZ_{m,n}$ & 0.485 & 0.052 & 0.312 & 0.368 & 0.248 & 0.212\tabularnewline
$HV_{m,n}$ & 0.717 & 0.613 & 0.207 & 0.300 & 0.123 & 0.040 & $HZ_{m,n}$ & 0.724 & 0.217 & 0.369 & 0.157 & 0.279 & 0.349\tabularnewline
$HE_{m,n}$ & 0.381 & 0.051 & 0.189 & 0.176 & 0.154 & 0.317 & $HA_{m,n}^{(1)}$ & 0.622 & 0.615 & 0.152 & 0.091 & 0.088 & 0.266\tabularnewline
$HE_{m,n}$ & 0.348 & 0.072 & 0.201 & 0.079 & 0.163 & 0.076 & $HA_{m,n}^{(1)}$ & 0.496 & 0.303 & 0.144 & 0.010 & 0.086 & 0.044\tabularnewline
$HE_{m,n}$ & 0.366 & 0.582 & 0.214 & 0.044 & 0.172 & 0.142 & $HA_{m,n}^{(1)}$ & 0.469 & 0.923 & 0.139 & 0.049 & 0.085 & 0.076\tabularnewline
$HE_{m,n}$ & 0.375 & 0.439 & 0.223 & 0.110 & 0.178 & 0.132 & $HA_{m,n}^{(1)}$ & 0.460 & 0.351 & 0.134 & 0.016 & 0.083 & 0.043\tabularnewline
$HE_{m,n}$ & 0.377 & 0.491 & 0.234 & 0.246 & 0.186 & 0.125 & $HA_{m,n}^{(1)}$ & 0.460 & 0.325 & 0.133 & 0.066 & 0.082 & 0.065\tabularnewline
$HE_{m,n}^{(1)}$ & 0.553 & 0.107 & 0.137 & 0.178 & 0.084 & 0.039 & $HA_{m,n}^{(2)}$ & 0.592 & 0.958 & 0.127 & 0.106 & 0.077 & 0.079\tabularnewline
$HE_{m,n}^{(1)}$ & 0.432 & 0.215 & 0.131 & 0.047 & 0.081 & 0.055 & $HA_{m,n}^{(2)}$ & 0.434 & 0.114 & 0.118 & 0.216 & 0.076 & 0.036\tabularnewline
$HE_{m,n}^{(1)}$ & 0.403 & 0.225 & 0.128 & 0.185 & 0.080 & 0.015 & $HA_{m,n}^{(2)}$ & 0.385 & 0.173 & 0.114 & 0.099 & 0.075 & 0.009\tabularnewline
$HE_{m,n}^{(1)}$ & 0.393 & 0.120 & 0.125 & 0.006 & 0.080 & 0.045 & $HA_{m,n}^{(2)}$ & 0.351 & 0.177 & 0.113 & 0.102 & 0.078 & 0.049\tabularnewline
$HE_{m,n}^{(1)}$ & 0.388 & 0.302 & 0.122 & 0.135 & 0.079 & 0.016 & $HA_{m,n}^{(2)}$ & 0.323 & 0.022 & 0.111 & 0.022 & 0.080 & 0.086\tabularnewline
$HE_{m,n}^{(2)}$ & 0.315 & 0.254 & 0.107 & 0.017 & 0.073 & 0.019 & $HA_{m,n}^{(3)}$ & 0.554 & 0.534 & 0.132 & 0.059 & 0.079 & 0.244\tabularnewline
$HE_{m,n}^{(2)}$ & 0.259 & 0.247 & 0.105 & 0.019 & 0.072 & 0.076 & $HA_{m,n}^{(3)}$ & 0.417 & 0.204 & 0.124 & 0.046 & 0.078 & 0.068\tabularnewline
$HE_{m,n}^{(2)}$ & 0.253 & 0.212 & 0.104 & 0.089 & 0.074 & 0.032 & $HA_{m,n}^{(3)}$ & 0.376 & 0.804 & 0.118 & 0.089 & 0.078 & 0.050\tabularnewline
$HE_{m,n}^{(2)}$ & 0.249 & 0.180 & 0.103 & 0.141 & 0.074 & 0.020 & $HA_{m,n}^{(3)}$ & 0.354 & 0.211 & 0.114 & 0.029 & 0.077 & 0.072\tabularnewline
$HE_{m,n}^{(2)}$ & - & - & 0.106 & 0.100 & 0.076 & 0.044 & $HA_{m,n}^{(3)}$ & 0.338 & 0.165 & 0.113 & 0.018 & 0.078 & 0.095\tabularnewline
$HC_{m,n}$ & 0.554 & 0.403 & 0.128 & 0.071 & 0.080 & 0.045 & $HB_{m,n}^{(1)}$ & - & - & 0.112 & 0.074 & 0.076 & 0.048\tabularnewline
$HC_{m,n}$ & 0.467 & 0.078 & 0.130 & 0.025 & 0.085 & 0.101 & $HB_{m,n}^{(1)}$ & 0.410 & 0.167 & 0.107 & 0.177 & 0.079 & 0.000\tabularnewline
$HC_{m,n}$ & 0.467 & 0.146 & 0.133 & 0.055 & 0.081 & 0.090 & $HB_{m,n}^{(1)}$ & 0.323 & 0.074 & 0.111 & 0.046 & 0.086 & 0.031\tabularnewline
$HC_{m,n}$ & 0.498 & 0.151 & 0.135 & 0.117 & 0.083 & 0.006 & $HB_{m,n}^{(1)}$ & 0.275 & 0.023 & 0.119 & 0.037 & 0.094 & 0.094\tabularnewline
$HC_{m,n}$ & 0.540 & 0.378 & 0.136 & 0.053 & 0.079 & 0.046 & $HB_{m,n}^{(1)}$ & 0.256 & 0.170 & 0.128 & 0.097 & 0.102 & 0.037\tabularnewline
$HW_{m,n}^{(1)}$ & 0.323 & 0.022 & 0.286 & 0.338 & 0.338 & 0.368 & $HB_{m,n}^{(2)}$ & - & - & 0.130 & 0.173 & 0.079 & 0.096\tabularnewline
$HW_{m,n}^{(1)}$ & 0.275 & 0.254 & 0.396 & 0.522 & 0.405 & 0.378 & $HB_{m,n}^{(2)}$ & 0.495 & 0.416 & 0.119 & 0.186 & 0.074 & 0.111\tabularnewline
$HW_{m,n}^{(1)}$ & 0.396 & 0.465 & 0.519 & 0.618 & 0.478 & 0.520 & $HB_{m,n}^{(2)}$ & 0.384 & 0.376 & 0.110 & 0.003 & 0.077 & 0.046\tabularnewline
$HW_{m,n}^{(1)}$ & 0.778 & 1.208 & 0.651 & 0.694 & 0.553 & 0.558 & $HB_{m,n}^{(2)}$ & 0.313 & 0.112 & 0.105 & 0.091 & 0.078 & 0.012\tabularnewline
$HW_{m,n}^{(1)}$ & 1.279 & 0.761 & 0.788 & 0.847 & 0.631 & 0.752 & $HB_{m,n}^{(2)}$ & 0.261 & 0.047 & 0.109 & 0.042 & 0.085 & 0.163\tabularnewline
$HP_{m,n}$ & 0.513 & 0.440 & 0.135 & 0.067 & 0.080 & 0.246 & $HJ_{m,n}$ & 0.489 & 0.039 & 0.111 & 0.098 & 0.086 & 0.111\tabularnewline
$HP_{m,n}$ & 0.410 & 0.214 & 0.126 & 0.041 & 0.078 & 0.067 & $HJ_{m,n}$ & 0.341 & 0.002 & 0.120 & 0.064 & 0.092 & 0.063\tabularnewline
$HP_{m,n}$ & 0.372 & 0.784 & 0.119 & 0.086 & 0.078 & 0.049 & $HJ_{m,n}$ & 0.290 & 0.374 & 0.124 & 0.063 & 0.097 & 0.086\tabularnewline
$HP_{m,n}$ & 0.348 & 0.207 & 0.114 & 0.028 & 0.077 & 0.072 & $HJ_{m,n}$ & 0.260 & 0.597 & 0.132 & 0.034 & 0.099 & 0.083\tabularnewline
$HP_{m,n}$ & 0.330 & 0.146 & 0.113 & 0.018 & 0.077 & 0.096 & $HJ_{m,n}$ & 0.260 & 0.198 & 0.137 & 0.179 & 0.125 & 0.082\tabularnewline
$HA_{m,n}$ & 0.555 & 0.913 & 0.110 & 0.062 & 0.077 & 0.032 & $HM_{m,n}$ & 1.244 & 0.942 & 0.127 & 0.059 & 0.088 & 0.162\tabularnewline
$HA_{m,n}$ & 0.369 & 0.203 & 0.108 & 0.163 & 0.081 & 0.037 & $HM_{m,n}$ & - & - & 0.119 & 0.147 & 0.088 & 0.124\tabularnewline
$HA_{m,n}$ & 0.303 & 0.039 & 0.113 & 0.037 & 0.089 & 0.071 & $HM_{m,n}$ & 0.281 & 0.464 & 0.117 & 0.104 & 0.089 & 0.174\tabularnewline
$HA_{m,n}$ & 0.265 & 0.001 & 0.122 & 0.030 & 0.097 & 0.017 & $HM_{m,n}$ & 0.272 & 0.418 & 0.115 & 0.109 & 0.090 & 0.130\tabularnewline
$HA_{m,n}$ & 0.255 & 0.202 & 0.133 & 0.028 & 0.105 & 0.059 & $HM_{m,n}$ & 0.282 & 0.105 & 0.115 & 0.058 & 0.091 & 0.004\tabularnewline
\hline 
\multicolumn{14}{c}{$R$ = RMSE, $B$ = $\left|Bias\right|$}\tabularnewline
\end{tabular}
\par\end{centering}

\end{table}

\begin{table}

\caption{RMSE and absolute bias of the spacing based Shannon entropy estimators
under the standard exponential distribution, $H(X)\thinspace=\thinspace1$}
\small
\centering{}%
\begin{tabular}{cccccccccccccc}
\hline 
\multirow{2}{*}{$H_{n}(X)$} & \multicolumn{2}{c}{$n=10\thinspace(3)$} & \multicolumn{2}{c}{$n=50\thinspace(7)$} & \multicolumn{2}{c}{$n=100\thinspace(10)$} & \multirow{2}{*}{$H_{n}(X)$} & \multicolumn{2}{c}{$n=10\thinspace(3)$} & \multicolumn{2}{c}{$n=50\thinspace(7)$} & \multicolumn{2}{c}{$n=100\thinspace(10)$}\tabularnewline
\cline{2-7} \cline{3-7} \cline{4-7} \cline{5-7} \cline{6-7} \cline{7-7} \cline{9-14} \cline{10-14} \cline{11-14} \cline{12-14} \cline{13-14} \cline{14-14} 
 & $R$ & $B$  & $R$ & $B$  & $R$ & $B$  &  & $R$ & $B$  & $R$ & $B$  & $R$ & $B$ \tabularnewline
\hline 
$HV_{m,n}$ & 0.677 & 0.765 & 0.199 & 0.298 & 0.131 & 0.013 & $HZ_{m,n}$ & 0.571 & 0.675 & 0.199 & 0.259 & 0.184 & 0.201\tabularnewline
$HV_{m,n}$ & 0.583 & 0.001 & 0.197 & 0.302 & 0.128 & 0.061 & $HZ_{m,n}$ & 0.369 & 0.390 & 0.255 & 0.003 & 0.215 & 0.159\tabularnewline
$HV_{m,n}$ & 0.566 & 0.967 & 0.197 & 0.050 & 0.130 & 0.196 & $HZ_{m,n}$ & 0.451 & 0.050 & 0.308 & 0.412 & 0.246 & 0.169\tabularnewline
$HV_{m,n}$ & 0.572 & 0.625 & 0.198 & 0.407 & 0.131 & 0.072 & $HZ_{m,n}$ & 0.685 & 0.541 & 0.368 & 0.450 & 0.278 & 0.286\tabularnewline
$HV_{m,n}$ & 0.599 & 0.381 & 0.199 & 0.023 & 0.129 & 0.057 & $HZ_{m,n}$ & 0.951 & 0.807 & 0.428 & 0.349 & 0.309 & 0.355\tabularnewline
$HE_{m,n}$ & 0.430 & 0.064 & 0.173 & 0.065 & 0.130 & 0.040 & $HA_{m,n}^{(1)}$ & 0.632 & 0.243 & 0.166 & 0.078 & 0.108  & 0.112\tabularnewline
$HE_{m,n}$ & 0.399 & 0.359 & 0.176 & 0.297 & 0.132 & 0.095 & $HA_{m,n}^{(1)}$ & 0.487 & 0.286 & 0.161 & 0.021 & 0.107 & 0.009\tabularnewline
$HE_{m,n}$ & 0.386 & 0.247 & 0.182 & 0.086 & 0.134 & 0.112 & $HA_{m,n}^{(1)}$ & 0.443 & 0.350 & 0.159 & 0.034 & 0.107 & 0.016\tabularnewline
$HE_{m,n}$ & 0.387 & 0.100 & 0.186 & 0.026 & 0.137 & 0.084 & $HA_{m,n}^{(1)}$ & 0.425 & 0.884 & 0.154 & 0.067 & 0.105 & 0.129\tabularnewline
$HE_{m,n}$ & 0.391 & 0.318 & 0.188 & 0.148 & 0.142 & 0.017 & $HA_{m,n}^{(1)}$ & 0.413 & 0.303 & 0.156 & 0.045 & 0.105 & 0.003\tabularnewline
$HE_{m,n}^{(1)}$ & 0.570 & 0.324 & 0.158 & 0.059 & 0.105 & 0.102 & $HA_{m,n}^{(2)}$ & 0.593 & 0.102 & 0.154 & 0.034 & 0.104 & 0.144\tabularnewline
$HE_{m,n}^{(1)}$ & 0.436 & 0.205 & 0.155 & 0.003 & 0.105 & 0.020 & $HA_{m,n}^{(2)}$ & 0.445 & 0.081 & 0.149 & 0.266 & 0.104 & 0.058\tabularnewline
$HE_{m,n}^{(1)}$ & 0.399 & 0.267 & 0.154 & 0.014 & 0.106 & 0.027 & $HA_{m,n}^{(2)}$ & 0.392 & 0.201 & 0.149 & 0.044 & 0.106 & 0.059\tabularnewline
$HE_{m,n}^{(1)}$ & 0.387 & 0.798 & 0.151 & 0.088 & 0.104 & 0.118 & $HA_{m,n}^{(2)}$ & 0.366 & 0.076 & 0.150 & 0.025 & 0.110 & 0.086\tabularnewline
$HE_{m,n}^{(1)}$ & 0.382 & 0.393 & 0.153 & 0.024 & 0.105 & 0.009 & $HA_{m,n}^{(2)}$ & 0.373 & 0.092 & 0.159 & 0.068 & 0.112 & 0.207\tabularnewline
$HE_{m,n}^{(2)}$ & 0.424 & 0.223 & 0.159 & 0.074 & 0.110 & 0.012 & $HA_{m,n}^{(3)}$ & 0.566 & 0.506 & 0.156 & 0.140 & 0.104 & 0.124\tabularnewline
$HE_{m,n}^{(2)}$ & 0.362 & 0.135 & 0.152 & 0.229 & 0.107 & 0.114 & $HA_{m,n}^{(3)}$ & 0.428 & 0.718 & 0.152 & 0.189 & 0.104 & 0.095\tabularnewline
$HE_{m,n}^{(2)}$ & 0.344 & 0.072 & 0.149 & 0.119 & 0.106 & 0.114 & $HA_{m,n}^{(3)}$ & 0.389 & 0.158 & 0.150 & 0.203 & 0.104 & 0.061\tabularnewline
$HE_{m,n}^{(2)}$ & 0.339 & 0.080 & 0.144 & 0.332 & 0.105 & 0.065 & $HA_{m,n}^{(3)}$ & 0.371 & 0.115 & 0.151 & 0.014 & 0.107 & 0.002\tabularnewline
$HE_{m,n}^{(2)}$ & - & - & 0.149 & 0.282 & 0.103 & 0.286 & $HA_{m,n}^{(3)}$ & 0.374 & 0.253 & 0.153 & 0.027 & 0.109 & 0.021\tabularnewline
$HC_{m,n}$ & 0.580 & 0.166 & 0.147 & 0.033 & 0.102 & 0.191 & $HB_{m,n}^{(1)}$ & - & - & 0.149 & 0.148 & 0.107 & 0.172\tabularnewline
$HC_{m,n}$ & 0.433 & 0.926 & 0.153 & 0.208 & 0.106 & 0.008 & $HB_{m,n}^{(1)}$ & 0.429 & 0.313 & 0.150 & 0.119 & 0.114 & 0.015\tabularnewline
$HC_{m,n}$ & 0.438 & 0.304 & 0.161 & 0.027 & 0.107 & 0.076 & $HB_{m,n}^{(1)}$ & 0.363 & 0.634 & 0.159 & 0.078 & 0.118 & 0.104\tabularnewline
$HC_{m,n}$ & 0.462 & 0.143 & 0.153 & 0.131 & 0.105 & 0.120 & $HB_{m,n}^{(1)}$ & 0.367 & 0.264 & 0.171 & 0.052 & 0.127 & 0.019\tabularnewline
$HC_{m,n}$ & 0.462 & 0.427 & 0.152 & 0.235 & 0.105 & 0.005 & $HB_{m,n}^{(1)}$ & 0.413 & 0.183 & 0.187 & 0.037 & 0.137 & 0.074\tabularnewline
$HW_{m,n}^{(1)}$ & 0.393 & 0.246 & 0.328 & 0.127 & 0.358 & 0.409 & $HB_{m,n}^{(2)}$ & - & - & 0.147 & 0.252 & 0.099 & 0.006\tabularnewline
$HW_{m,n}^{(1)}$ & 0.370 & 0.498 & 0.440 & 0.243 & 0.429 & 0.432 & $HB_{m,n}^{(2)}$ & 0.451 & 0.389 & 0.146 & 0.173 & 0.101 & 0.052\tabularnewline
$HW_{m,n}^{(1)}$ & 0.555 & 0.109 & 0.564 & 0.724 & 0.499 & 0.371 & $HB_{m,n}^{(2)}$ & 0.353 & 0.113 & 0.144 & 0.064 & 0.109 & 0.089\tabularnewline
$HW_{m,n}^{(1)}$ & 0.976 & 0.726 & 0.699 & 0.404 & 0.576 & 0.570 & $HB_{m,n}^{(2)}$ & 0.340 & 0.321 & 0.153 & 0.017 & 0.119 & 0.049\tabularnewline
$HW_{m,n}^{(1)}$ & 1.504 & 1.543 & 0.843 & 0.934 & 0.656 & 0.665 & $HB_{m,n}^{(2)}$ & 0.348 & 0.020 & 0.168 & 0.025 & 0.122 & 0.088\tabularnewline
$HP_{m,n}$ & 0.488 & 0.603 & 0.151 & 0.142 & 0.104 & 0.046 & $HJ_{m,n}$ & 0.503 & 0.395 & 0.156 & 0.063 & 0.117 & 0.208\tabularnewline
$HP_{m,n}$ & 0.405 & 0.432 & 0.151 & 0.223 & 0.106 & 0.023 & $HJ_{m,n}$ & 0.3600 & 0.079 & 0.166 & 0.096 & 0.127 & 0.245\tabularnewline
$HP_{m,n}$ & 0.373 & 0.350 & 0.152 & 0.141 & 0.108 & 0.031 & $HJ_{m,n}$ & 0.389 & 0.315 & 0.177 & 0.224 & 0.171 & 0.213\tabularnewline
$HP_{m,n}$ & 0.370 & 0.106 & 0.156 & 0.137 & 0.111 & 0.070 & $HJ_{m,n}$ & 0.378 & 0.445 & 0.182 & 0.202 & 0.190 & 0.223\tabularnewline
$HP_{m,n}$ & 0.382 & 0.038 & 0.160 & 0.019 & 0.115 & 0.101 & $HJ_{m,n}$ & 0.438 & 0.157 & 0.206 & 0.073 & 0.199 & 0.219\tabularnewline
$HA_{m,n}$ & 0.569 & 0.627 & 0.148 & 0.160 & 0.109 & 0.098 & $HM_{m,n}$ & 1.228 & 1.377 & 0.175 & 0.462 & 0.121 & 0.231\tabularnewline
$HA_{m,n}$ & 0.407 & 0.278 & 0.153 & 0.135 & 0.114 & 0.064 & $HM_{m,n}$ & - & - & 0.168 & 0.165 & 0.121 & 0.086\tabularnewline
$HA_{m,n}$ & 0.361 & 0.551 & 0.162 & 0.244 & 0.121 & 0.057 & $HM_{m,n}$ & 0.399 & 0.645 & 0.165 & 0.357 & 0.121 & 0.143\tabularnewline
$HA_{m,n}$ & 0.376 & 0.071 & 0.176 & 0.186 & 0.130 & 0.080 & $HM_{m,n}$ & 0.369 & 0.441 & 0.166 & 0.050 & 0.124 & 0.040\tabularnewline
$HA_{m,n}$ & 0.437 & 0.313 & 0.196 & 0.227 & 0.140 & 0.110 & $HM_{m,n}$ & 0.373 & 0.136 & 0.173 & 0.105 & 0.126 & 0.146\tabularnewline
\hline 
\multicolumn{14}{c}{$R$ = RMSE, $B$ = $\left|Bias\right|$}\tabularnewline
\end{tabular}
\end{table}

\begin{table}

\caption{RMSE and absolute bias of the kernel density estimation based Shannon
entropy estimators under the uniform, standard normal and standard
exponential distributions, $n$ = 10}

\centering{}%
\begin{tabular}{ccccccc}
\hline 
$H_{n}(X)$ & \multicolumn{2}{c}{$U(0,1)$} & \multicolumn{2}{c}{$N(0,1)$} & \multicolumn{2}{c}{$Exp(1)$}\tabularnewline
\hline 
 & RMSE & $\left|Bias\right|$ & RMSE & $\left|Bias\right|$ & RMSE & $\left|Bias\right|$\tabularnewline
\hline 
$HAL_{d=1,n}$ & 0.1865 & 0.1904 & 0.2812 & 0.3500 & 0.3800 & 0.4327\tabularnewline
$HAL_{d=2,n}$ & 0.2783  & 0.1193 & 0.5932 & 0.5202 & 0.5598 & 0.1127\tabularnewline
$HAL_{d=3,n}$ & 0.5663  & 0.0868 & 1.1695 & 0.3132 & 0.7355 & 0.8772\tabularnewline
$HAL_{d=5,n}$ & 1.9423  & 1.9359 & 3.0566 & 2.4248 & 1.7063 & 0.4496\tabularnewline
$HAN_{m,n}$ & 0.1887 & 0.1723 & 0.2958 & 0.8231 & 0.3749 & 0.0661\tabularnewline
$HAN_{m,n}$ & 0.1900 & 0.2264 & 0.2896 & 0.0989 & 0.3860 & 0.3886\tabularnewline
$HAN_{m,n}$ & 0.1963 & 0.2635 & 0.2672 & 0.2064 & 0.3874 & 0.0600\tabularnewline
$HAN_{m,n}$ & 0.2137 & 0.1886 & 0.2556 & 0.1801 & 0.4080 & 0.1721\tabularnewline
$HAN_{m,n}$ & 0.2506 & 0.2083 & 0.2785 & 0.0366 & 0.4117 & 0.5502\tabularnewline
$HZA_{m,n}^{(1)}$ & 0.2862 & 0.0400 & 0.3469 & 0.1874 & 0.5365 & 0.1151\tabularnewline
$HZA_{m,n}^{(1)}$ & 0.3923 & 0.2962 & 0.3396 & 0.0065 & 0.7217 & 1.2074\tabularnewline
$HZA_{m,n}^{(1)}$ & 0.3981 & 0.3928 & 0.4678 & 0.2787 & 0.6841 & 0.2981\tabularnewline
$HZA_{m,n}^{(1)}$ & 0.4128 & 0.2853 & 0.6263 & 0.7908 & 0.5867 & 0.4878\tabularnewline
$HZA_{m,n}^{(1)}$ & 0.4992 & 0.6822 & 0.7379 & 1.2288 & 0.5580 & 0.2631\tabularnewline
$HZA_{m,n}^{(2)}$ & 0.4738 & 0.0681 & 1.0088 & 0.2302 & 1.4629 & 1.6898\tabularnewline
$HZA_{m,n}^{(2)}$ & 0.7054 & 0.6813 & 1.0332 & 0.3830 & 1.7826 & 2.0100\tabularnewline
$HZA_{m,n}^{(2)}$ & 0.6911 & 0.5591 & 0.7182 & 0.7049 & 1.5462 & 0.5360\tabularnewline
$HZA_{m,n}^{(2)}$ & 0.6640 & 0.4981 & 0.5677 & 0.0111 & 1.1109 & 0.9835\tabularnewline
$HZA_{m,n}^{(2)}$ & 0.7297 & 0.9945 & 0.6341 & 1.1218 & 0.8411 & 1.4178\tabularnewline
$HAN_{m,n}^{2}$ & 0.2556 & 0.3902 & 0.4278 & 0.8321 & 0.4334 & 0.6882\tabularnewline
$HAN_{m,n}^{2}$ & 0.1767 & 0.0785 & 0.3131 & 0.0759 & 0.3642 & 0.2476\tabularnewline
$HAN_{m,n}^{2}$ & 0.1785 & 0.1957 & 0.2860 & 0.0716 & 0.3724 & 0.2705\tabularnewline
$HAN_{m,n}^{2}$ & 0.1852 & 0.0946 & 0.2704 & 0.6479 & 0.4153 & 0.0229\tabularnewline
$HAN_{m,n}^{2}$ & 0.1872 & 0.0097 & 0.2819 & 0.8970 & 0.3986 & 0.0932\tabularnewline
\hline 
\end{tabular}
\end{table}

\begin{table}

\caption{RMSE and absolute bias of the kernel density estimation based Shannon
entropy estimators under the uniform, standard normal and standard
exponential distributions, $n$ = 50}

\centering{}%
\begin{tabular}{ccccccc}
\hline 
$H_{n}(X)$ & \multicolumn{2}{c}{$U(0,1)$} & \multicolumn{2}{c}{$N(0,1)$} & \multicolumn{2}{c}{$Exp(1)$}\tabularnewline
\hline 
 & RMSE & $\left|Bias\right|$ & RMSE & $\left|Bias\right|$ & RMSE & $\left|Bias\right|$\tabularnewline
\hline 
$HAL_{d=1,n}$ & 0.1156 & 0.0668 & 0.1113 & 0.1077 & 0.2232 & 0.3315\tabularnewline
$HAL_{d=2,n}$ & 0.1804 & 0.0324 & 0.2341 & 0.3687 & 0.3437 & 0.2490\tabularnewline
$HAL_{d=3,n}$ & 0.1422  & 0.1135 & 0.4993 & 0.7184 & 0.4123  & 0.6147\tabularnewline
$HAL_{d=5,n}$ & 0.6033 & 0.4978 & 1.5845 & 1.2290 & 0.3891  & 0.3470\tabularnewline
$HAN_{m,n}$ & 0.1122 & 0.1103 & 0.1210 & 0.1106 & 0.2100 & 0.2053\tabularnewline
$HAN_{m,n}$ & 0.1147 & 0.1069 & 0.1212 & 0.1294 & 0.2078 & 0.2714\tabularnewline
$HAN_{m,n}$ & 0.1150 & 0.1029 & 0.1200 & 0.1259 & 0.2044 & 0.0636\tabularnewline
$HAN_{m,n}$ & 0.1155 & 0.1669 & 0.1222 & 0.0230 & 0.2036 & 0.1233\tabularnewline
$HAN_{m,n}$ & 0.1172 & 0.0981 & 0.1174 & 0.0743 & 0.1862 & 0.3036\tabularnewline
$HAN_{m,n}$ & 0.1223 & 0.1200 & 0.1195 & 0.0776 & 0.1911 & 0.0383\tabularnewline
$HAN_{m,n}$ & 0.1273 & 0.1262 & 0.1130 & 0.1571 & 0.1947 & 0.1468\tabularnewline
$HAN_{m,n}$ & 0.1312 & 0.0670 & 0.1155 & 0.0832 & 0.1825 & 0.1650\tabularnewline
$HAN_{m,n}$ & 0.1352 & 0.1606 & 0.1102 & 0.0682 & 0.1849 & 0.2386\tabularnewline
$HZA_{m,n}^{(1)}$ & 0.1271 & 0.1740 & 0.1214 & 0.0053 & 0.2573 & 0.2790\tabularnewline
$HZA_{m,n}^{(1)}$ & 0.4074 & 0.3335 & 0.2527 & 0.2084 & 0.5425 & 0.6266\tabularnewline
$HZA_{m,n}^{(1)}$ & 0.4697 & 0.5892 & 0.2963 & 0.3051 & 0.6168 & 0.5495\tabularnewline
$HZA_{m,n}^{(1)}$ & 0.4671 & 0.4487 & 0.2759 & 0.2923 & 0.6372 & 0.8438\tabularnewline
$HZA_{m,n}^{(1)}$ & 0.4358 & 0.3231 & 0.2237 & 0.2738 & 0.6093 & 0.8670\tabularnewline
$HZA_{m,n}^{(1)}$ & 0.3733 & 0.3331 & 0.1614 & 0.0156 & 0.5830 & 0.7221\tabularnewline
$HZA_{m,n}^{(1)}$ & 0.3067 & 0.2813 & 0.1348 & 0.1186 & 0.5017 & 0.3879\tabularnewline
$HZA_{m,n}^{(1)}$ & 0.2419 & 0.2134 & 0.1701 & 0.0397 & 0.4364 & 0.2781\tabularnewline
$HZA_{m,n}^{(1)}$ & 0.1690 & 0.1451 & 0.2530 & 0.0348 & 0.3648 & 0.7934\tabularnewline
$HZA_{m,n}^{(2)}$ & 0.2246 & 0.2888 & 1.2506 & 1.2813 & 1.3389 & 1.3914\tabularnewline
$HZA_{m,n}^{(2)}$ & 0.7807 & 0.5795 & 1.7513 & 2.2409 & 1.9107 & 1.3229\tabularnewline
$HZA_{m,n}^{(2)}$ & 0.9078 & 0.7056 & 1.8270 & 1.8738 & 2.0809 & 2.0782\tabularnewline
$HZA_{m,n}^{(2)}$ & 0.9008 & 0.9375 & 1.7427 & 1.9888 & 2.1012 & 2.0271\tabularnewline
$HZA_{m,n}^{(2)}$ & 0.8451 & 0.7430 & 1.6084 & 1.6408 & 2.0229 & 2.5751\tabularnewline
$HZA_{m,n}^{(2)}$ & 0.7446 & 0.7490 & 1.4482 & 1.8820 & 1.9173 & 1.4768\tabularnewline
$HZA_{m,n}^{(2)}$ & 0.6291 & 0.6352 & 1.2460 & 1.4963 & 1.7787 & 1.9972\tabularnewline
$HZA_{m,n}^{(2)}$ & 0.4636 & 0.3658 & 1.0361 & 0.9181 & 1.6546 & 0.9012\tabularnewline
$HZA_{m,n}^{(2)}$ & 0.3445 & 0.2121 & 0.8075 & 0.6014 & 1.4616 & 0.6535\tabularnewline
$HAN_{m,n}^{2}$ & 0.2548 & 0.2006 & 0.3154 & 0.3134 & 0.3132 & 0.1795\tabularnewline
$HAN_{m,n}^{2}$ & 0.1024 & 0.0512 & 0.1805 & 0.0351 & 0.1864 & 0.1467\tabularnewline
$HAN_{m,n}^{2}$ & 0.0554 & 0.1133 & 0.1540 & 0.0933 & 0.1585 & 0.0129\tabularnewline
$HAN_{m,n}^{2}$ & 0.0528 & 0.0026 & 0.1361 & 0.2230 & 0.1520 & 0.1334\tabularnewline
$HAN_{m,n}^{2}$ & 0.0605 & 0.0020 & 0.1269 & 0.0439 & 0.1548 & 0.1345\tabularnewline
$HAN_{m,n}^{2}$ & 0.0756 & 0.0610 & 0.1258 & 0.1226 & 0.1622 & 0.0087\tabularnewline
$HAN_{m,n}^{2}$ & 0.0870 & 0.0154 & 0.1137 & 0.0963 & 0.1670 & 0.0161\tabularnewline
$HAN_{m,n}^{2}$ & 0.0948 & 0.0918 & 0.1093 & 0.1641 & 0.1840 & 0.3566\tabularnewline
$HAN_{m,n}^{2}$ & 0.1001 & 0.1775 & 0.1125 & 0.1096 & 0.2000  & 0.1205\tabularnewline
\hline 
\end{tabular}
\end{table}

\begin{table}

\caption{RMSE and absolute bias of the kernel density estimation based Shannon
entropy estimators under the uniform, standard normal and standard
exponential distributions, $n$ = 100}
\small
\centering{}%
\begin{tabular}{ccccccc}
\hline 
$H_{n}(X)$ & \multicolumn{2}{c}{$U(0,1)$} & \multicolumn{2}{c}{$N(0,1)$} & \multicolumn{2}{c}{$Exp(1)$}\tabularnewline
\hline 
 & RMSE & $\left|Bias\right|$ & RMSE & $\left|Bias\right|$ & RMSE & $\left|Bias\right|$\tabularnewline
\hline 
$HAL_{d=1,n}$ & 0.1036 & 0.0558 & 0.0801 & 0.0936 & 0.1915 & 0.0955\tabularnewline
$HAL_{d=2,n}$ & 0.1831 & 0.2019 & 0.1662  & 0.1075 & 0.3119  & 0.3131\tabularnewline
$HAL_{d=3,n}$ & 0.1852 & 0.0528 & 0.3845  & 0.2919 & 0.3988 & 0.0628\tabularnewline
$HAL_{d=5,n}$ & 0.3541 & 0.3131 & 1.2851 & 1.1510 & 0.3861 & 0.4715\tabularnewline
$HAN_{m,n}$ & 0.1002 & 0.0788 & 0.0793 & 0.0635 & 0.1808 & 0.1267\tabularnewline
$HAN_{m,n}$ & 0.1041 & 0.1477 & 0.0821 & 0.0247 & 0.1828 & 0.0938\tabularnewline
$HAN_{m,n}$ & 0.1035 & 0.1429 & 0.0800 & 0.0043 & 0.1743 & 0.1060\tabularnewline
$HAN_{m,n}$ & 0.1028 & 0.1396 & 0.0827 & 0.0391 & 0.1707 & 0.0768\tabularnewline
$HAN_{m,n}$ & 0.1035 & 0.1185 & 0.0874 & 0.0634 & 0.1661 & 0.2015\tabularnewline
$HAN_{m,n}$ & 0.1054 & 0.0851 & 0.0877 & 0.0863 & 0.1649 & 0.0399\tabularnewline
$HAN_{m,n}$ & 0.1061 & 0.1068 & 0.0876 & 0.1728 & 0.1591 & 0.1072\tabularnewline
$HAN_{m,n}$ & 0.1070 & 0.0477 & 0.0862 & 0.0501 & 0.1571 & 0.0816\tabularnewline
$HAN_{m,n}$ & 0.1102 & 0.1244 & 0.0847 & 0.0645 & 0.1553 & 0.0327\tabularnewline
$HAN_{m,n}$ & 0.1102 & 0.0943 & 0.0846 & 0.0346 & 0.1535 & 0.1013\tabularnewline
$HAN_{m,n}$ & 0.1146 & 0.0917 & 0.0857 & 0.2433 & 0.1526 & 0.2831\tabularnewline
$HAN_{m,n}$ & 0.1162 & 0.1382 & 0.0837 & 0.0217 & 0.1496 & 0.0960\tabularnewline
$HZA_{m,n}^{(1)}$ & 0.1083 & 0.0152 & 0.0800 & 0.0371 & 0.2008 & 0.4317\tabularnewline
$HZA_{m,n}^{(1)}$ & 0.4147 & 0.4316 & 0.2889 & 0.4400 & 0.5094 & 0.4478\tabularnewline
$HZA_{m,n}^{(1)}$ & 0.5036 & 0.4658 & 0.3641 & 0.3040 & 0.6126 & 0.8127\tabularnewline
$HZA_{m,n}^{(1)}$ & 0.5300 & 0.4414 & 0.3802 & 0.3505 & 0.6460 & 0.7230\tabularnewline
$HZA_{m,n}^{(1)}$ & 0.5223 & 0.5344 & 0.3657 & 0.3375 & 0.6470 & 0.6442\tabularnewline
$HZA_{m,n}^{(1)}$ & 0.5101 & 0.4748 & 0.3376 & 0.3914 & 0.6367 & 0.5471\tabularnewline
$HZA_{m,n}^{(1)}$ & 0.4812 & 0.4817 & 0.2962 & 0.3821 & 0.6172 & 0.6498\tabularnewline
$HZA_{m,n}^{(1)}$ & 0.4464 & 0.4836 & 0.2535 & 0.2763 & 0.5841 & 0.6060\tabularnewline
$HZA_{m,n}^{(1)}$ & 0.4054 & 0.4978 & 0.1952 & 0.1427 & 0.5516 & 0.4385\tabularnewline
$HZA_{m,n}^{(1)}$ & 0.3559 & 0.2583 & 0.1577 & 0.0964 & 0.5133 & 0.1148\tabularnewline
$HZA_{m,n}^{(1)}$ & 0.3106 & 0.2103 & 0.1117 & 0.0053 & 0.4738 & 0.6900\tabularnewline
$HZA_{m,n}^{(1)}$ & 0.2572 & 0.1862 & 0.0926 & 0.0035 & 0.4209 & 0.6187\tabularnewline
$HZA_{m,n}^{(2)}$ & 0.2058 & 0.2109 & 1.3124 & 1.3471 & 1.3082 & 1.2562\tabularnewline
$HZA_{m,n}^{(2)}$ & 0.8148 & 0.8083 & 1.8902 & 1.8034 & 1.9459 & 1.7213\tabularnewline
$HZA_{m,n}^{(2)}$ & 0.9858 & 1.0117 & 2.0409 & 1.9147 & 2.1253 & 1.8749\tabularnewline
$HZA_{m,n}^{(2)}$ & 1.0479 & 1.0629 & 2.0531 & 2.2247 & 2.1985 & 2.0943\tabularnewline
$HZA_{m,n}^{(2)}$ & 1.0447 & 0.9805 & 2.0188 & 2.3502 & 2.1828 & 2.2810\tabularnewline
$HZA_{m,n}^{(2)}$ & 1.0164 & 0.9256 & 1.9552 & 1.8141 & 2.1631 & 2.2344\tabularnewline
$HZA_{m,n}^{(2)}$ & 0.9610 & 1.0706 & 1.8663 & 2.0072 & 2.1043 & 1.8393\tabularnewline
$HZA_{m,n}^{(2)}$ & 0.8972 & 0.9328 & 1.7696 & 1.5065 & 2.0404 & 2.2977\tabularnewline
$HZA_{m,n}^{(2)}$ & 0.8276 & 0.8998 & 1.6542 & 1.8115 & 1.9633 & 1.8080\tabularnewline
$HZA_{m,n}^{(2)}$ & 0.7379 & 0.7995 & 1.5412 & 1.4589 & 1.8927 & 1.5388\tabularnewline
$HZA_{m,n}^{(2)}$ & 0.6433 & 0.4870 & 1.4014 & 1.6309 & 1.8058 & 1.3577\tabularnewline
$HZA_{m,n}^{(2)}$ & 0.5539 & 0.6287 & 1.2724 & 1.4225 & 1.6971 & 2.0027\tabularnewline
$HAN_{m,n}^{2}$ & 0.2600 & 0.2855 & 0.2939 & 0.2653 & 0.2886 & 0.1716\tabularnewline
$HAN_{m,n}^{2}$ & 0.1086 & 0.0950 & 0.1628 & 0.1462 & 0.1618 & 0.3316\tabularnewline
$HAN_{m,n}^{2}$ & 0.0596 & 0.0790 & 0.1223 & 0.2844 & 0.1193 & 0.1695\tabularnewline
$HAN_{m,n}^{2}$ & 0.0348 & 0.0037 & 0.1011 & 0.1433 & 0.1073 & 0.0025\tabularnewline
$HAN_{m,n}^{2}$ & 0.0302 & 0.0301 & 0.0937 & 0.1273 & 0.1011 & 0.0688\tabularnewline
$HAN_{m,n}^{2}$ & 0.0359 & 0.0523 & 0.0892 & 0.1189 & 0.1011 & 0.0509\tabularnewline
$HAN_{m,n}^{2}$ & 0.0442 & 0.0508 & 0.0851 & 0.0125 & 0.1101 & 0.0005\tabularnewline
$HAN_{m,n}^{2}$ & 0.0515 & 0.0223 & 0.0821 & 0.0893 & 0.1118 & 0.2115\tabularnewline
$HAN_{m,n}^{2}$ & 0.0600 & 0.0805 & 0.0831 & 0.0952 & 0.1192 & 0.0372\tabularnewline
$HAN_{m,n}^{2}$ & 0.0656 & 0.0070 & 0.0792 & 0.0281 & 0.1266 & 0.0956\tabularnewline
$HAN_{m,n}^{2}$ & 0.0735 & 0.0653 & 0.0758 & 0.1063 & 0.1336 & 0.1983\tabularnewline
$HAN_{m,n}^{2}$ & 0.0775 & 0.1089 & 0.0763 & 0.1490 & 0.1391 & 0.0391\tabularnewline
\hline 
\end{tabular}
\end{table}

\begin{table}

\caption{RMSE and absolute bias of the $k-$nearest neighbour based Shannon
entropy estimators under the uniform distribution, $d=1,2,3,5$; $n$
= 10, 50, 100}

\centering{}%
\begin{tabular}{cccccccc}
\hline 
$d$ & $H_{n}(X)$ & \multicolumn{2}{c}{$n=10$} & \multicolumn{2}{c}{$n=50$} & \multicolumn{2}{c}{$n=100$}\tabularnewline
\hline 
 &  & RMSE & $\left|Bias\right|$ & RMSE & $\left|Bias\right|$ & RMSE & $\left|Bias\right|$\tabularnewline
\hline 
\multirow{12}{*}{$d=1$} & $HKL$ & 0.4690 & 0.8470 & 0.2060 & 0.2963 & 0.1469 & 0.1213\tabularnewline
 & $HV$ & 0.8167 & 1.6547 & 0.5278 & 0.0051 & 0.4841 & 0.2577\tabularnewline
 & $HS_{k=1}$ & 0.4693 & 0.8528 & 0.2053 & 0.0580 & 0.1421 & 0.0370\tabularnewline
 & $HS_{k=3}$ & 1.5799 & 1.4664 & 1.5162 & 1.6120 & 1.5118 & 1.4383\tabularnewline
 & $HS_{k=5}$ & 2.2317 & 2.4866 & 2.1151 & 2.1232 & 2.0949 & 2.0563\tabularnewline
 & $HN$ & 0.6880 & 0.4819 & 0.3522 & 0.1065 & 0.2885 & 0.2853\tabularnewline
 & $HK_{k=1}$ & 1.4838 & 1.4121 & 1.3995 & 1.6380 & 1.3914 & 1.4634\tabularnewline
 & $HK_{k=3}$ & 1.5303 & 1.6739 & 1.4181 & 1.3118 & 1.3984 & 1.2607\tabularnewline
 & $HK_{k=5}$ & 1.5920 & 2.1252 & 1.4239 & 1.2879 & 1.4038 & 1.4005\tabularnewline
 & $HL_{k=1}$ & 0.4659 & 1.2766 & 0.2050 & 0.2894 & 0.1471 & 0.1535\tabularnewline
 & $HL_{k=3}$ & 0.2917 & 0.0041 & 0.1126 & 0.0340 & 0.0796 & 0.0654\tabularnewline
 & $HL_{k=5}$ & 0.2908 & 0.0808 & 0.0946 & 0.0082 & 0.0667 & 0.0498\tabularnewline
\hline 
\multirow{12}{*}{$d=2$} & $HKL$ & 6.4251 & 6.3092 & 7.7421 & 7.7589 & 8.3632 & 8.3655\tabularnewline
 & $HV$ & 0.7749 & 0.7118 & 0.4431 & 0.0301 & 0.4055 & 0.5215\tabularnewline
 & $HS_{k=1}$ & 0.5413 & 0.8728 & 0.2357 & 0.0584 & 0.1684 & 0.0927\tabularnewline
 & $HS_{k=3}$ & 1.2998 & 1.3804 & 1.6558 & 1.5883 & 1.6096 & 1.4062\tabularnewline
 & $HS_{k=5}$ & 1.8660 & 1.8834 & 2.2956 & 2.4160 & 2.2256 & 2.1625\tabularnewline
 & $HN$ & 0.7346 & 0.3256 & 0.3735 & 0.0819 & 0.3147 & 0.6352\tabularnewline
 & $HK_{k=1}$ & 3.0371 & 2.6507 & 2.8659 & 2.6312 & 2.8344 & 2.9363\tabularnewline
 & $HK_{k=3}$ & 3.1933 & 3.1909 & 2.9402 & 2.9663 & 2.8830 & 2.8275\tabularnewline
 & $HK_{k=5}$ & 3.3187 & 2.9257 & 2.8681 & 2.8146 & 2.9221 & 2.9273\tabularnewline
 & $HL_{k=1}$ & 0.5430 & 0.9971 & 0.2342 & 0.2979 & 0.1654 & 0.0869\tabularnewline
 & $HL_{k=3}$ & 0.4920 & 0.3612 & 0.2036 & 0.3635 & 0.1371 & 0.2085\tabularnewline
 & $HL_{k=5}$ & 0.5819 & 0.4946 & 0.2316 & 0.1421 & 0.1590 & 0.1663\tabularnewline
\hline 
\multirow{12}{*}{$d=3$} & $HKL$ & 7.6282 & 7.8478 & 10.8663 & 10.9023 & 12.2329 & 12.7121\tabularnewline
 & $HV$ & 0.8402 & 0.4164 & 0.3491 & 0.0665 & 0.2917 & 0.5153\tabularnewline
 & $HS_{k=1}$ & 0.6733 & 0.5828 & 0.3127 & 0.2536 & 0.2386 & 0.6570\tabularnewline
 & $HS_{k=3}$ & 2.1852 & 2.2749 & 1.8749 & 1.9794 & 1.7913 & 1.9091\tabularnewline
 & $HS_{k=5}$ & 2.9603 & 2.3488 & 2.5434 & 2.5543 & 2.4385 & 2.5141\tabularnewline
 & $HN$ & 0.8892 & 0.6475 & 0.3274 & 0.0417 & 0.2462 & 0.6088\tabularnewline
 & $HK_{k=1}$ & 4.5971 & 4.8019 & 4.3904 & 4.4228 & 4.3436 & 4.8210\tabularnewline
 & $HK_{k=3}$ & 4.8859 & 4.9941 & 4.5408 & 4.6487 & 4.4537 & 4.5730\tabularnewline
 & $HK_{k=5}$ & 5.0974 & 4.9819 & 4.6279 & 4.6402 & 4.5184 & 4.5948\tabularnewline
 & $HL_{k=1}$ & 0.6762 & 0.5885 & 0.3128 & 0.2538 & 0.2387 & 0.6570\tabularnewline
 & $HL_{k=3}$ & 0.7592 & 0.7806 & 0.3971 & 0.4796 & 0.3047 & 0.4091\tabularnewline
 & $HL_{k=5}$ & 0.9415 & 0.7685 & 0.4740 & 0.4712 & 0.3640 & 0.4309\tabularnewline
\hline 
\multirow{12}{*}{$d=5$} & $HKL$ & 16.3756 & 16.8167 & 22.8715 & 22.7615 & 25.5985 & 25.7500\tabularnewline
 & $HV$ & 1.3123 & 1.5646 & 0.6487 & 0.3620 & 0.4785 & 0.6150\tabularnewline
 & $HS_{k=1}$ & 1.1334 & 1.3787 & 0.7123 & 0.5506 & 0.6035 & 0.7261\tabularnewline
 & $HS_{k=3}$ & 2.9345 & 2.6750 & 2.4246 & 2.5288 & 2.2894 & 2.4756\tabularnewline
 & $HS_{k=5}$ & 3.8183 & 3.4442 & 3.1466 & 3.1737 & 2.9883 & 3.1037\tabularnewline
 & $HN$ & 1.4579 & 1.7645 & 0.7117 & 0.4390 & 0.5312 & 0.6773\tabularnewline
 & $HK_{k=1}$ & 7.9433 & 8.3704 & 7.6056 & 7.4924 & 7.5128 & 7.6627\tabularnewline
 & $HK_{k=3}$ & 8.4017 & 8.1668 & 7.8617 & 7.9706 & 7.7237 & 7.9121\tabularnewline
 & $HK_{k=5}$ & 8.7135 & 8.3526 & 8.0023 & 8.0322 & 7.8403 & 7.9569\tabularnewline
 & $HL_{k=1}$ & 1.1381 & 1.3844 & 0.7125 & 0.5508 & 0.6036 & 0.7261\tabularnewline
 & $HL_{k=3}$ & 1.4789 & 1.1808 & 0.9359 & 1.0290 & 0.7954 & 0.9756\tabularnewline
 & $HL_{k=5}$ & 1.7685 & 1.3666 & 1.0727 & 1.0906 & 0.9097 & 1.0204\tabularnewline
\hline 
\end{tabular}
\end{table}

\begin{table}

\caption{RMSE and absolute bias of the $k-$nearest neighbour based Shannon
entropy estimators under the standard normal distribution, $d=1,2,3,5$,
$n$ = 10, 50, 100}

\centering{}%
\begin{tabular}{cccccccc}
\hline 
$d$ & $H_{n}(X)$ & \multicolumn{2}{c}{$n=10$} & \multicolumn{2}{c}{$n=50$} & \multicolumn{2}{c}{$n=100$}\tabularnewline
\hline 
 &  & RMSE & $\left|Bias\right|$ & RMSE & $\left|Bias\right|$ & RMSE & $\left|Bias\right|$\tabularnewline
\hline 
\multirow{12}{*}{$d=1$} & $HKL$ & 0.4927 & 0.2724 & 0.2254 & 0.0321 & 0.1688 & 0.0643\tabularnewline
 & $HV$ & 0.7234 & 0.1975 & 0.3656 & 0.2418 & 0.3097 & 0.2882\tabularnewline
 & $HS_{k=1}$ & 0.4931 & 0.2781 & 0.2255 & 0.0319 & 0.1688 & 0.0642\tabularnewline
 & $HS_{k=3}$ & 1.4111 & 1.0259 & 1.4719 & 1.3982 & 1.4861 & 1.4146\tabularnewline
 & $HS_{k=5}$ & 1.9854 & 1.4127 & 2.0398 & 2.0295 & 2.0592 & 2.0443\tabularnewline
 & $HN$ & 0.8777 & 0.1759 & 0.5303 & 0.4923 & 0.4917 & 0.5241\tabularnewline
 & $HK_{k=1}$ & 1.4887 & 1.1684 & 1.3964 & 1.4286 & 1.3987 & 1.4556\tabularnewline
 & $HK_{k=3}$ & 1.3592 & 0.9724 & 1.3692 & 1.2949 & 1.3778 & 1.3060\tabularnewline
 & $HK_{k=5}$ & 1.3597 & 0.7759 & 1.3552 & 1.3429 & 1.3684 & 1.3524\tabularnewline
 & $HL_{k=1}$ & 0.4927 & 0.2724 & 0.2254 & 0.0321 & 0.1688 & 0.0643\tabularnewline
 & $HL_{k=3}$ & 0.3457 & 0.4684 & 0.1501 & 0.1016 & 0.1087 & 0.0854\tabularnewline
 & $HL_{k=5}$ & 0.3262 & 0.6649 & 0.1380 & 0.0536 & 0.0990 & 0.0390\tabularnewline
\hline 
\multirow{12}{*}{$d=2$} & $HKL$ & 3.2820 & 2.5110 & 4.9989  & 5.2487 & 5.7131 & 5.7478\tabularnewline
 & $HV$ & 1.0124  & 0.3751 & 0.8482 & 1.1297 & 0.8241 & 0.8352\tabularnewline
 & $HS_{k=1}$ & 0.5328  & 0.8367 & 0.2657  & 0.2119 & 0.1851 & 0.0079\tabularnewline
 & $HS_{k=3}$ & 1.4584  & 0.4307 & 1.4307  & 1.4256 & 1.4444  & 1.7704\tabularnewline
 & $HS_{k=5}$ & 2.1003  & 1.2895 & 1.9974  & 2.0031 & 2.0110  & 2.2926\tabularnewline
 & $HN$ & 1.2089  & 0.1062 & 0.9811  & 1.2758 & 0.9496  & 0.9666\tabularnewline
 & $HK_{k=1}$ & 2.7749 & 1.9962 & 2.7507  & 2.9949 & 2.7540 & 2.7856\tabularnewline
 & $HK_{k=3}$ & 2.7641  & 1.7636 & 2.7082  & 2.7086 & 2.7195 & 3.0481\tabularnewline
 & $HK_{k=5}$ & 2.8406  & 2.0390 & 2.6952 & 2.7027 & 2.7044  & 2.9870\tabularnewline
 & $HL_{k=1}$ & 0.5317 & 0.8309 & 0.2657 & 0.2121 & 0.1851  & 0.0080\tabularnewline
 & $HL_{k=3}$ & 0.4132  & 1.0635 & 0.1991 & 0.0742 & 0.1384  & 0.2705\tabularnewline
 & $HL_{k=5}$ & 0.3823  & 0.7881 & 0.1915 & 0.0801 & 0.1398  & 0.2094\tabularnewline
\hline 
\multirow{12}{*}{$d=3$} & $HKL$ & 7.2023  & 6.2410 & 10.5799 & 10.6734 & 11.9948 & 12.4489\tabularnewline
 & $HV$ & 1.5791  & 0.0173 & 1.4037 & 1.4877 & 1.3932  & 2.0199\tabularnewline
 & $HS_{k=1}$ & 0.5906  & 1.0240 & 0.2811  & 0.0247 & 0.2111 & 0.3937\tabularnewline
 & $HS_{k=3}$ & 1.5686 & 1.3752 & 1.4158  & 1.5310 & 1.4228  & 1.7520\tabularnewline
 & $HS_{k=5}$ & 2.2914 & 2.0472 & 1.9984 & 2.1581 & 1.9955 & 2.3120\tabularnewline
 & $HN$ & 1.7788  & 0.2138 & 1.5079  & 1.5959 & 1.4849 & 2.1135\tabularnewline
 & $HK_{k=1}$ & 4.1738  & 3.1951 & 4.1060  & 4.1939 & 4.1070 & 4.5577\tabularnewline
 & $HK_{k=3}$ & 4.2402 & 4.0944 & 4.0756  & 4.2002 & 4.0817 & 4.4160\tabularnewline
 & $HK_{k=5}$ & 4.4056  & 4.1830 & 4.0796 & 4.2440 & 4.0737  & 4.3927\tabularnewline
 & $HL_{k=1}$ & 0.5898 & 1.0183 & 0.2811  & 0.0249 & 0.2111  & 0.3938\tabularnewline
 & $HL_{k=3}$ & 0.4779  & 0.1190 & 0.2238 & 0.0312 & 0.1712 & 0.2521\tabularnewline
 & $HL_{k=5}$ & 0.4811 & 0.0304 & 0.2133  & 0.0750 & 0.1673 & 0.2287\tabularnewline
\hline 
\multirow{12}{*}{$d=5$} & $HKL$ & 15.4875  & 13.7993 & 22.1517 & 22.3543 & 24.9585  & 25.1748\tabularnewline
 & $HV$ & 2.9386  & 0.3521 & 2.6578  & 2.9153 & 2.6342  & 2.9257\tabularnewline
 & $HS_{k=1}$ & 0.6894  & 1.6388 & 0.3163  & 0.1434 & 0.2451  & 0.1508\tabularnewline
 & $HS_{k=3}$ & 1.9043  & 1.7177 & 1.5053  & 1.4836 & 1.4650  & 1.8211\tabularnewline
 & $HS_{k=5}$ & 2.7758  & 3.3017 & 2.1424 & 2.0493 & 2.0720  & 2.4627\tabularnewline
 & $HN$ & 3.1274 & 0.5520 & 2.7337  & 2.9923 & 2.6960  & 2.9881\tabularnewline
 & $HK_{k=1}$ & 7.0596 & 5.3530 & 6.8875  & 7.0853 & 6.8741 & 7.0874\tabularnewline
 & $HK_{k=3}$ & 7.3259 & 7.2094 & 6.9302 & 6.9254 & 6.8921  & 7.2576\tabularnewline
 & $HK_{k=5}$ & 7.6483 & 8.2102 & 6.9912  & 6.9078 & 6.9197 & 7.3159\tabularnewline
 & $HL_{k=1}$ & 0.6897  & 1.6330 & 0.3163  & 0.1436 & 0.2451  & 0.1509\tabularnewline
 & $HL_{k=3}$ & 0.6702 & 0.2234 & 0.2549  & 0.0162 & 0.1931  & 0.3211\tabularnewline
 & $HL_{k=5}$ & 0.8491  & 1.2242 & 0.2484  & 0.0339 & 0.1814 & 0.3794\tabularnewline
\hline 
\end{tabular}
\end{table}

\begin{table}

\caption{RMSE and absolute bias of the $k-$nearest neighbour based Shannon
entropy estimators under the standard exponential distribution, $d=1,2,3,5$,
$n$ = 10, 50, 100}

\centering{}%
\begin{tabular}{cccccccc}
\hline 
$d$ & $H_{n}(X)$ & \multicolumn{2}{c}{$n=10$} & \multicolumn{2}{c}{$n=50$} & \multicolumn{2}{c}{$n=100$}\tabularnewline
\hline 
 &  & RMSE & $\left|Bias\right|$ & RMSE & $\left|Bias\right|$ & RMSE & $\left|Bias\right|$\tabularnewline
\hline 
\multirow{12}{*}{$d=1$} & $HKL$ & 0.5564 & 0.3710 & 0.2466 & 0.0637 & 0.1719 & 0.0030\tabularnewline
 & $HV$ & 0.8018 & 0.5453 & 0.3559 & 0.0819 & 0.2484 & 0.0057\tabularnewline
 & $HS_{k=1}$ & 0.5571 & 0.3653 & 0.2466 & 0.0639 & 0.1725 & 0.1816\tabularnewline
 & $HS_{k=3}$ & 1.4900 & 1.5564 & 1.4935 & 1.4551 & 1.4914 & 1.4662\tabularnewline
 & $HS_{k=5}$ & 2.0964 & 2.0289 & 2.0755 & 2.0643 & 2.0758 & 1.9314\tabularnewline
 & $HN$ & 0.8638 & 0.7071 & 0.4312 & 0.6659 & 0.3436 & 0.3371\tabularnewline
 & $HK_{k=1}$ & 1.5013 & 1.6652 & 1.4178 & 1.3327 & 1.4041 & 1.3883\tabularnewline
 & $HK_{k=3}$ & 1.4564 & 1.3454 & 1.3970 & 1.3693 & 1.3939 & 1.3698\tabularnewline
 & $HK_{k=5}$ & 1.4793 & 1.2070 & 1.3921 & 1.3776 & 1.3899 & 1.3649\tabularnewline
 & $HL_{k=1}$ & 0.5592 & 0.2244 & 0.2466 & 0.0637 & 0.1719 & 0.0030\tabularnewline
 & $HL_{k=3}$ & 0.4055 & 0.0954 & 0.1739 & 0.0272 & 0.1259 & 0.0215\tabularnewline
 & $HL_{k=5}$ & 0.3935 & 0.2338 & 0.1657 & 0.0189 & 0.1163 & 0.0264\tabularnewline
\hline 
\multirow{12}{*}{$d=2$} & $HKL$ & 3.5058 & 3.6179 & 5.1205 & 4.9991 & 5.7865  & 5.7393\tabularnewline
 & $HV$ & 1.1393 & 0.8509 & 0.7081  & 0.3987 & 0.5935 & 0.4519\tabularnewline
 & $HS_{k=1}$ & 0.6798 & 0.2702 & 0.3086 & 0.0376 & 0.2092 & 0.0006\tabularnewline
 & $HS_{k=3}$ & 1.8335 & 1.7030 & 1.6291 & 1.3785 & 1.5855 & 1.4660\tabularnewline
 & $HS_{k=5}$ & 2.5425 & 2.4909 & 2.2421 & 2.0132 & 2.1861 & 2.1625\tabularnewline
 & $HN$ & 1.3001 & 1.1199 & 0.8287 & 0.5448 & 0.71055  & 0.5834\tabularnewline
 & $HK_{k=1}$ & 3.0022 & 3.1031 & 2.8735 & 2.7453 & 2.8281 & 2.7771\tabularnewline
 & $HK_{k=3}$ & 3.1250 & 3.0359 & 2.9043 & 2.6615 & 2.8593 & 2.7437\tabularnewline
 & $HK_{k=5}$ & 3.2762 & 3.2404 & 2.9389  & 2.712851 & 2.8791 & 2.8569\tabularnewline
 & $HL_{k=1}$ & 0.6806 & 0.2760 & 0.3087 & 0.0374 & 0.2092  & 0.0006\tabularnewline
 & $HL_{k=3}$ & 0.6379 & 0.2088 & 0.2637 & 0.1213 & 0.1818  & 0.0340\tabularnewline
 & $HL_{k=5}$ & 0.7096 & 0.4133 & 0.2726 & 0.0699 & 0.1850  & 0.0792\tabularnewline
\hline 
\multirow{12}{*}{$d=3$} & $HKL$ & 7.6629 & 8.1375 & 10.8810  & 10.5450 & 12.2333 & 12.3581\tabularnewline
 & $HV$ & 1.8088 & 2.1624 & 1.3188 & 0.7462 & 1.1965 & 1.3327\tabularnewline
 & $HS_{k=1}$ & 0.8547  & 0.8724 & 0.4102  & 0.1036 & 0.2899  & 0.3030\tabularnewline
 & $HS_{k=3}$ & 2.2647  & 2.6021 & 1.8709  & 1.6614 & 1.7840  & 1.7960\tabularnewline
 & $HS_{k=5}$ & 3.0782 & 3.0616 & 2.5257 & 2.4461 & 2.4214 & 2.4701\tabularnewline
 & $HN$ & 1.9955 & 2.3935 & 1.4197  & 0.8544 & 1.2866 & 1.4262\tabularnewline
 & $HK_{k=1}$ & 4.6429 & 5.0916 & 4.4095  & 4.0656 & 4.3461 & 4.4670\tabularnewline
 & $HK_{k=3}$ & 4.9244 & 5.3212 & 4.5265  & 4.3307 & 4.4417  & 4.4599\tabularnewline
 & $HK_{k=5}$ & 5.1819 & 5.1974 & 4.6042  & 4.5320 & 4.4986 & 4.5507\tabularnewline
 & $HL_{k=1}$ & 0.8571 & 0.8782 & 0.4103 & 0.1034 & 0.2899 & 0.3031\tabularnewline
 & $HL_{k=3}$ & 0.9580 & 1.1078 & 0.4551 & 0.1616 & 0.3345 & 0.2960\tabularnewline
 & $HL_{k=5}$ & 1.1516 & 0.9840 & 0.5136 & 0.3630 & 0.3818 & 0.3868\tabularnewline
\hline 
\multirow{12}{*}{$d=5$} & $HKL$ & 16.5903  & 17.3355 & 22.9704  & 22.5229 & 25.6710  & 25.6593\tabularnewline
 & $HV$ & 3.6703 & 4.5265 & 2.9321 & 2.2312 & 2.7425  & 2.6975\tabularnewline
 & $HS_{k=1}$ & 1.4636  & 1.8974 & 0.8602  & 0.3120 & 0.7045 & 0.6353\tabularnewline
 & $HS_{k=3}$ & 3.2772 & 3.7460 & 2.5837 & 2.2774 & 2.4093 & 2.5708\tabularnewline
 & $HS_{k=5}$ & 4.2862  & 4.7487 & 3.3164  & 3.1113 & 3.1160 & 3.2770\tabularnewline
 & $HN$ & 3.8569 & 4.7264 & 3.0075  & 2.3082 & 2.8042  & 2.7598\tabularnewline
 & $HK_{k=1}$ & 8.1713 & 8.8892 & 7.7087  & 7.2539 & 7.5873  & 7.5719\tabularnewline
 & $HK_{k=3}$ & 8.6906 & 9.2378 & 8.0084 & 7.7193 & 7.8367  & 8.0074\tabularnewline
 & $HK_{k=5}$ & 9.1439 & 9.6571 & 8.1632  & 7.9699 & 7.9630 & 8.1302\tabularnewline
 & $HL_{k=1}$ & 1.4680 & 1.9032 & 0.8604 & 0.3122 & 0.7045 & 0.6354\tabularnewline
 & $HL_{k=3}$ & 1.8844  & 2.2518 & 1.1185  & 0.7776 & 0.9310  & 1.0709\tabularnewline
 & $HL_{k=5}$ & 2.2960  & 2.6711 & 1.2661  & 1.0282 & 1.0529  & 1.1937\tabularnewline
\hline 
\end{tabular}
\end{table}

\begin{sidewaystable}

\caption{Estimators with least RMSE values among those with reduced bias values
across the considered distributions}

\centering{}%
\begin{tabular}{cccc}
\hline 
$n$ & \multicolumn{3}{c}{Distributions}\tabularnewline
\hline 
 & Uniform & Normal & Exponential\tabularnewline
\hline 
10 & $HE_{m,n}^{(2)}$; $HB_{m,n}^{(2)}$; $HA_{m,n}$; $HA_{m,n}^{(2)}$;
$HM_{m,n}^{(2)}$ & $HE_{m,n}^{(2)}$; $HA_{m,n}$; $HB_{m,n}^{(1)}$; $HB_{m,n}^{(2)}$;
$HJ_{m,n}$ & $HA_{m,n}$; $HE_{m,n}^{(2)}$; $HB_{m,n}^{(1)}$; $HB_{m,n}^{(2)}$;
$HJ_{m,n}$\tabularnewline
50 & $HP_{m,n}$; $HA_{m,n}^{(2)}$; $HA_{m,n}^{(3)}$; $HA_{m,n}$; $HB_{m,n}^{(1)}$ & $HE_{m,n}^{(2)}$; $HA_{m,n}$; $HA_{m,n}^{(2)}$; $HB_{m,n}^{(1)}$;
$HB_{m,n}^{(2)}$ & $HE_{m,n}^{(2)}$; $HC_{m,n}$; $HA_{m,n}$; $HB_{m,n}^{(1)}$; $HB_{m,n}^{(2)}$\tabularnewline
100 & $HP_{m,n}$; $HA_{m,n}^{(2)}$; $HA_{m,n}^{(3)}$; $HB_{m,n}^{(2)}$;
$HA_{m,n}$ & $HE_{m,n}^{(2)}$; $HB_{m,n}^{(2)}$; $HB_{m,n}^{(1)}$; $HP_{m,n}$;
$HA_{m,n}^{(3)}$ & $HC_{m,n}$; $HP_{m,n}$; $HA_{m,n}^{(2)}$; $HA_{m,n}^{(3)}$; $HB_{m,n}^{(2)}$\tabularnewline
\hline 
\end{tabular}
\end{sidewaystable}

Further look at the Tables 1 - 3 as well as Table 10 showed that for
the uniform distribution, $HB_{m,n}^{(2)}$, $HA_{m,n}$, $HA_{m,n}^{(2)}$,
$HA_{m,n}^{(3)}$, and $HP_{m,n}$ generally recorded best performance
results irrespective of the sample size. Also, $HE_{m,n}^{(2)}$,
$HA_{m,n}$, $HB_{m,n}^{(1)}$, $HB_{m,n}^{(2)}$, and $HA_{m,n}^{(2)}$
turned out to generally give the best results for the standard normal
distribution and for the standard exponential distribution, the $HA_{m,n}$,
$HA_{m,n}^{(2)}$, $HB_{m,n}^{(1)}$, $HB_{m,n}^{(2)}$, and $HC_{m,n}$
were considered the best estimators at all the $n$ considered. Based
on this, one may suggest that irrespective of the distribution under
consideration and $n$ of interest, the $HA_{m,n}$, $HA_{m,n}^{(2)}$,
$HB_{m,n}^{(1)}$, and $HB_{m,n}^{(2)}$ are expected to give a good
estimation of the Shannon differential entropy. Again, while $HE_{m,n}^{(2)}$,
$HB_{m,n}^{(2)}$, $HA_{m,n}$, $HB_{m,n}^{(1)}$, and $HJ_{m,n}$
are generally good estimators at small sample sizes, $HP_{m,n}$,
$HA_{m,n}^{(2)}$, $HA_{m,n}^{(3)}$, $HB_{m,n}^{(2)}$, and $HA_{m,n}$
are generally good estimators at large sample sizes.

From Table 4 - 6, it is shown that the estimators have varying performances
of the RMSE and absolute bias across sample sizes and distributions
considered. With the exception of the $HAL_{d=1},$the remainder of
the estimators, which are functions of $m$, have their performances
dependent on $m$, even within a sample size and a distribution.
Also, as expected, all the estimators have decreasing least RMSE and
$\left|bias\right|$ as $n$ increased in all the distributions considered.
At $d=1$, unlike the spacings based counterparts, the kernel density
estimation (KDE) based estimators have the most efficient ones the
same across all the distributions considered. Precisely, the $HAL_{d=1},$$HAN_{m,n}$
and $HAN_{m,n}^{2}$ recorded least RMSE and $\left|bias\right|$
values in all the distributions and all the sample sizes considered,
except at $n=100$, where $HZA_{m,n}^{(1)}$ was among the best three
estimators but only under the standard normal distribution, with RMSE
of 0.0800, a little lower than that of the $HAL_{d=1},$which was
0.0801. In terms of the $\left|bias\right|$, while $HAL_{d=1}$ recorded
0.0936, the $HZA_{m,n}^{(1)}$ stood at 0.0371. This notwithstanding,
since $HAL_{d=1}$ far outperformed it under all the distributions
at $n=10,\thinspace50$ and under the uniform and standard exponential
distributions at $n=100$, it is reasonable to conclude that the $HAL_{d=1},$$HAN_{m,n}$
and $HAN_{m,n}^{2}$ were generally the most efficient among the KDE
class of estimators considered.

Again, Tables 7 - 9 showed that all the six estimators considered
in the class of $k-$nearest neighbour ($k$NN) estimators have decreasing
RMSE and $\left|bias\right|$ values as $n$ increased in all the
distributions and variable dimensions ($d$) considered, except the
$HKL$, which maintained it only at $d=1$. At $d=1$, the estimators
with most efficient values were the $HKL$, $HS_{k=1}$ and $HL_{k=5}$
in all the $n$ and distributions considered. At $d>1,\thinspace(d=2,\thinspace3,\thinspace5)$,
the $HV$, $HS_{k=1}$ and $HL$ were observed to be the best estimators
across the $n$ and distributions considered, except for the uniform
distribution, where $HN$ outperformed the $HL$ especially at large
$n=50,\thinspace100$ and $d=5$ and also outperformed $HV$ especially
at large $n$ and $d=2,\thinspace3$. It is also of importance to
mention that at $d>1$, the $HKL$ estimator recorded the worst performances
among all the $k$NN class of estimators considered.

In the $k$NN class of estimators, the number of nearest neighbours,
($k$), was varied at $k=1,\thinspace3,\thinspace5$, for the three
estimators, $HS$, $HK$, and $HL$, which are functions of $k$.
It is observed that while the $HS$ maintained least RMSE and $\left|bias\right|$
values at $k=1$, in all the $n$, $d$, and distributions considered,
the $HL$ maintained least RMSE and $\left|bias\right|$ values at
$k=5$ in all the $n$ and distributions considered at $d=1$, but
generally moved to $k=1$ as $d$ increased.

In the light of the foregoing, the performances of the best estimators
across the three classes considered are presented in Tables 11 and
12 respectively for $d=1$ and $d=2,\thinspace3,\thinspace5$.

\begin{table}

\caption{The RMSE and $\left|bias\right|$ values of the best estimators across
the classes under the three distributions, $n=10,\thinspace50,\thinspace100$,
$d=1$}

\centering{}%
\begin{tabular}{cccccccc}
\hline 
\multirow{3}{*}{$H_{n}(x)$} & \multirow{3}{*}{$n$} & \multicolumn{6}{c}{Distributions}\tabularnewline
\cline{3-8} \cline{4-8} \cline{5-8} \cline{6-8} \cline{7-8} \cline{8-8} 
 &  & \multicolumn{2}{c}{Uniform} & \multicolumn{2}{c}{Normal} & \multicolumn{2}{c}{Exponential}\tabularnewline
\cline{3-8} \cline{4-8} \cline{5-8} \cline{6-8} \cline{7-8} \cline{8-8} 
 &  & RMSE & $\left|bias\right|$ & RMSE & $\left|bias\right|$ & RMSE & $\left|bias\right|$\tabularnewline
$HA_{m,n}$ & 10 & 0.1622 & 0.1680 & 0.2554 & 0.2016 & 0.3612 & 0.5510\tabularnewline
$HA_{m,n}^{(2)}$  & 10 & 0.1658 & 0.1387 & 0.3229 & 0.0216 & 0.3659 & 0.0762\tabularnewline
$HB_{m,n}^{(1)}$ & 10 & 0.1722 & 0.1795 & 0.2558 & 0.1702 & 0.3631 & 0.6343\tabularnewline
$HB_{m,n}^{(2)}$ & 10 & 0.1451 & 0.0197 & 0.2605 & 0.0466 & 0.3403 & 0.3212\tabularnewline
$HAL_{d=1}$ & 10 & 0.1865 & 0.1904 & 0.2812 & 0.3500 & 0.3800 & 0.4327\tabularnewline
$HAN_{m,n}$  & 10 & 0.1887 & 0.1723 & 0.2556 & 0.1801 & 0.3749 & 0.0661\tabularnewline
$HAN_{m,n}^{2}$  & 10 & 0.1767 & 0.0785 & 0.2704 & 0.6479 & 0.3642 & 0.2476\tabularnewline
$HKL$ & 10 & 0.4690 & 0.8470 & 0.4927 & 0.2724 & 0.5564 & 0.3710\tabularnewline
$HS_{k=1}$ & 10 & 0.4693 & 0.8528 & 0.4931 & 0.2781 & 0.5571 & 0.3653\tabularnewline
$HL_{k=5}$ & 10 & 0.2908 & 0.0808 & 0.3262 & 0.6649 & 0.3935 & 0.2338\tabularnewline
 &  &  &  &  &  &  & \tabularnewline
$HA_{m,n}$ & 50 & 0.0393 & 0.0594 & 0.1075 & 0.1625 & 0.1484 & 0.1598\tabularnewline
$HA_{m,n}^{(2)}$  & 50 & 0.0362 & 0.1058 & 0.1110 & 0.0222 & 0.1488 & 0.0437\tabularnewline
$HB_{m,n}^{(1)}$ & 50 & 0.0403 & 0.0375 & 0.1074 & 0.1765 & 0.1487 & 0.1477\tabularnewline
$HB_{m,n}^{(2)}$ & 50 & 0.0405 & 0.0140 & 0.1046 & 0.0914 & 0.1436 & 0.0635\tabularnewline
$HAL_{d=1}$ & 50 & 0.1156 & 0.0668 & 0.1113 & 0.1077 & 0.2232 & 0.3315\tabularnewline
$HAN_{m,n}$  & 50 & 0.1122 & 0.1103 & 0.1102 & 0.0682 & 0.1825 & 0.1650\tabularnewline
$HAN_{m,n}^{2}$  & 50 & 0.0528 & 0.0026 & 0.1093 & 0.1641 & 0.1520 & 0.1334\tabularnewline
$HKL$ & 50 & 0.2060 & 0.2963 & 0.2254 & 0.0321 & 0.2466 & 0.0637\tabularnewline
$HS_{k=1}$ & 50 & 0.2053 & 0.0580 & 0.2255 & 0.0319 & 0.2466 & 0.0639\tabularnewline
$HL_{k=5}$ & 50 & 0.0946 & 0.0082 & 0.1380 & 0.0536 & 0.1657 & 0.0189\tabularnewline
 &  &  &  &  &  &  & \tabularnewline
$HA_{m,n}$ & 100 & 0.0263 & 0.0040 & 0.0768 & 0.0323 & 0.1097 & 0.0983\tabularnewline
$HA_{m,n}^{(2)}$  & 100 & 0.0197 & 0.0294 & 0.0754 & 0.0089 & 0.1038 & 0.0579\tabularnewline
$HB_{m,n}^{(1)}$ & 100 & 0.0241 & 0.0103 & 0.0763 & 0.0478 & 0.1069 & 0.1716\tabularnewline
$HB_{m,n}^{(2)}$ & 100 & 0.0247 & 0.0184 & 0.0738 & 0.1107 & 0.0995 & 0.0063\tabularnewline
$HAL_{d=1}$ & 100 & 0.1036 & 0.0558 & 0.0801 & 0.0936 & 0.1915 & 0.0955\tabularnewline
$HAN_{m,n}$  & 100 & 0.1002 & 0.0788 & 0.0793 & 0.0635 & 0.1496 & 0.0960\tabularnewline
$HAN_{m,n}^{2}$  & 100 & 0.0302 & 0.0301 & 0.0758 & 0.1063 & 0.1011 & 0.0688\tabularnewline
$HKL$ & 100 & 0.1469 & 0.1213 & 0.1688 & 0.0643 & 0.1719 & 0.0030\tabularnewline
$HS_{k=1}$ & 100 & 0.1421 & 0.0370 & 0.1688 & 0.0642 & 0.1725 & 0.1816\tabularnewline
$HL_{k=5}$ & 100 & 0.0667 & 0.0498 & 0.0990 & 0.0390 & 0.1163 & 0.0264\tabularnewline
\hline 
\end{tabular}
\end{table}

\begin{table}

\caption{The RMSE and $\left|bias\right|$ values of the best estimators across
the classes under the three distributions, $n=10,\thinspace50,\thinspace100$,
$d=2,\thinspace3,\thinspace5$}

\centering{}%
\begin{tabular}{ccccccccc}
\hline 
\multirow{3}{*}{$d$} & \multirow{3}{*}{$H_{n}(X)$} & \multirow{3}{*}{$n$} & \multicolumn{6}{c}{Distributions}\tabularnewline
\cline{4-9} \cline{5-9} \cline{6-9} \cline{7-9} \cline{8-9} \cline{9-9} 
 &  &  & \multicolumn{2}{c}{Uniform} & \multicolumn{2}{c}{Normal} & \multicolumn{2}{c}{Exponential}\tabularnewline
\cline{4-9} \cline{5-9} \cline{6-9} \cline{7-9} \cline{8-9} \cline{9-9} 
 &  &  & RMSE & $\left|bias\right|$ & RMSE & $\left|bias\right|$ & RMSE & $\left|bias\right|$\tabularnewline
\multirow{12}{*}{2} & $HAL_{d=2,n}$ & 10 & 0.2783  & 0.1193 & 0.5932 & 0.5202 & 0.5598 & 0.1127\tabularnewline
 & $HV$ & 10 & 0.7749 & 0.7118 & 1.0124  & 0.3751 & 1.1393 & 0.8509\tabularnewline
 & $HS_{k=1,n}$ & 10 & 0.5413 & 0.8728 & 0.5328  & 0.8367 & 0.6798 & 0.2702\tabularnewline
 & $HL$ & 10 & 0.4920 & 0.3612 & 0.3823  & 0.7881 & 0.6379 & 0.2088\tabularnewline
 & $HAL_{d=2,n}$ & 50 & 0.1804 & 0.0324 & 0.2341 & 0.3687 & 0.3437 & 0.2490\tabularnewline
 & $HV$ & 50 & 0.4431 & 0.0301 & 0.8482 & 1.1297 & 0.7081  & 0.3987\tabularnewline
 & $HS_{k=1,n}$ & 50 & 0.2357 & 0.0584 & 0.2657  & 0.2119 & 0.3086 & 0.0376\tabularnewline
 & $HL$ & 50 & 0.2036 & 0.3635 & 0.1915 & 0.0801 & 0.2637 & 0.1213\tabularnewline
 & $HAL_{d=2,n}$ & 100 & 0.1831 & 0.2019 & 0.1662  & 0.1075 & 0.3119  & 0.3131\tabularnewline
 & $HV$ & 100 & 0.4055 & 0.5215 & 0.8241 & 0.8352 & 0.5935 & 0.4519\tabularnewline
 & $HS_{k=1,n}$ & 100 & 0.1684 & 0.0927 & 0.1851 & 0.0079 & 0.2092 & 0.0006\tabularnewline
 & $HL$ & 100 & 0.1371 & 0.2085 & 0.1384  & 0.2705 & 0.1818  & 0.0340\tabularnewline
 &  &  &  &  &  &  &  & \tabularnewline
\multirow{12}{*}{3} & $HAL_{d=3,n}$ & 10 & 0.5663  & 0.0868 & 1.1695 & 0.3132 & 0.7355 & 0.8772\tabularnewline
 & $HV$ & 10 & 0.8402 & 0.4164 & 1.5791  & 0.0173 & 1.8088 & 2.1624\tabularnewline
 & $HS_{k=1}$ & 10 & 0.6733 & 0.5828 & 0.5906  & 1.0240 & 0.8547  & 0.8724\tabularnewline
 & $HL$ & 10 & 0.6762 & 0.5885 & 0.4779  & 0.1190 & 0.8571 & 0.8782\tabularnewline
 & $HAL_{d=3,n}$ & 50 & 0.1422  & 0.1135 & 0.4993 & 0.7184 & 0.4123  & 0.6147\tabularnewline
 & $HV$ & 50 & 0.3491 & 0.0665 & 1.4037 & 1.4877 & 1.3188 & 0.7462\tabularnewline
 & $HS_{k=1}$ & 50 & 0.3127 & 0.2536 & 0.2811  & 0.0247 & 0.4102  & 0.1036\tabularnewline
 & $HL$ & 50 & 0.3128 & 0.2538 & 0.2133  & 0.0750 & 0.4103 & 0.1034\tabularnewline
 & $HAL_{d=3,n}$ & 100 & 0.1852 & 0.0528 & 0.3845  & 0.2919 & 0.3988 & 0.0628\tabularnewline
 & $HV$ & 100 & 0.2917 & 0.5153 & 1.3932  & 2.0199 & 1.1965 & 1.3327\tabularnewline
 & $HS_{k=1}$ & 100 & 0.2386 & 0.6570 & 0.2111 & 0.3937 & 0.2899  & 0.3030\tabularnewline
 & $HL$ & 100 & 0.2387 & 0.6570 & 0.1673 & 0.2287 & 0.2899 & 0.3031\tabularnewline
 &  &  &  &  &  &  &  & \tabularnewline
\multirow{12}{*}{5} & $HAL_{d=5,n}$ & 10 & 1.9423  & 1.9359 & 3.0566 & 2.4248 & 1.7063 & 0.4496\tabularnewline
 & $HV$ & 10 & 2.9386  & 0.3521 & 2.9386  & 0.3521 & 3.6703 & 4.5265\tabularnewline
 & $HS_{k=1}$ & 10 & 0.6894  & 1.6388 & 0.6894  & 1.6388 & 1.4636  & 1.8974\tabularnewline
 & $HL$ & 10 & 1.1381 & 1.3844 & 0.6702 & 0.2234 & 1.4680 & 1.9032\tabularnewline
 & $HAL_{d=5,n}$ & 50 & 0.6033 & 0.4978 & 1.5845 & 1.2290 & 0.3891  & 0.3470\tabularnewline
 & $HV$ & 50 & 1.5646 & 0.6487 & 2.6578  & 2.9153 & 2.9321 & 2.2312\tabularnewline
 & $HS_{k=1}$ & 50 & 1.3787 & 0.7123 & 0.3163  & 0.1434 & 0.8602  & 0.3120\tabularnewline
 & $HL$ & 50 & 0.7125 & 0.5508 & 0.2484  & 0.0339 & 0.8604 & 0.3122\tabularnewline
 & $HAL_{d=5,n}$ & 100 & 0.3541 & 0.3131 & 1.2851 & 1.1510 & 0.3861 & 0.4715\tabularnewline
 & $HV$ & 100 & 0.3620 & 0.4785 & 2.6342  & 2.9257 & 2.7425  & 2.6975\tabularnewline
 & $HS_{k=1}$ & 100 & 0.5506 & 0.6035 & 0.2451  & 0.1508 & 0.7045 & 0.6353\tabularnewline
 & $HL$ & 100 & 0.6036 & 0.7261 & 0.1814 & 0.3794 & 0.7045 & 0.6354\tabularnewline
\hline 
\end{tabular}
\end{table}

From Table 11, it can be seen that at $d=1$, the four best spacings
based estimators outperformed the rest of best estimators in all the
sample sizes and distributions considered. This was however with the
exception of the $HAN_{m,n}^{2}$, a spacings based KDE estimator,
which was almost at par with the outperforming ones. Also, the three
best $k$NN based estimators occupied the rear in terms of both RMSE
and $\left|bias\right|$ values. However, the $HL_{k=5}$, a $k$NN
based estimator, outperformed all the other estimators in this class
in all $n$ and distributions considered. As a result, this study
has identified four best estimators of the Shannon differential entropy,
all from the spacings class, namely: the $HA_{m,n}$, $HA_{m,n}^{(2)}$,
$HB_{m,n}^{(1)}$ and $HB_{m,n}^{(2)}$ with $HAN_{m,n}^{2}$ as the
best from the KDE class and the $HL_{k=5}$ as the best from the $k$NN
class of estimators. The six are presented in this study under the
univariate sphere ($d=1$) with a pseudo name 'best estimators'.

In order to further study the asymptotic distributions and unbiasedness
of the estimators, the boxplots of the univariate best estimators
were plotted at $n=10,\thinspace50,\thinspace100$ respectively under
the three distributions considered. For each estimator, a total of
1000 estimates were obtained at each sample size of each distribution
and the boxplots of the 1000 estimates in each case were plotted and
presented in Figures 1, 2, and 3 respectively for the uniform, standard
normal and standard exponential distributions. In each boxplot, the
red horizontal dashed line represents the true population entropy
of the distribution.

From Figures 1 - 3, the significant shrinkage of the whole 1000 estimates,
with more visibility in the middle 50 percentiles, towards the medians
with increasing sample size in all the estimators is observed. This
is indicative of two important asymptotic properties of the estimators.
First, the biases of the estimators vanished asymptotically, according
to Biau and Devroye (2015). Secondly, the asymptotic variances of
the estimators tend to the true population variances. Also, there
were continuous median shifts of the estimates obtained from all the
estimators in all the distributions considered from $n=10$ to $n=$
100. Furthermore, all the estimators do not have the same asymptotic
distributions in all the three parent distributions. While the $HA_{m,n}^{(2)}$
and $HL_{k=5}$ maintained asymptotic symmetric distributions in all
the three parent distributions, the rest of the best estimators maintained
asymptotic skewed distributions with various degrees and types (left/right
skewed) of skewness across the parent distributions.

\begin{figure}
\includegraphics{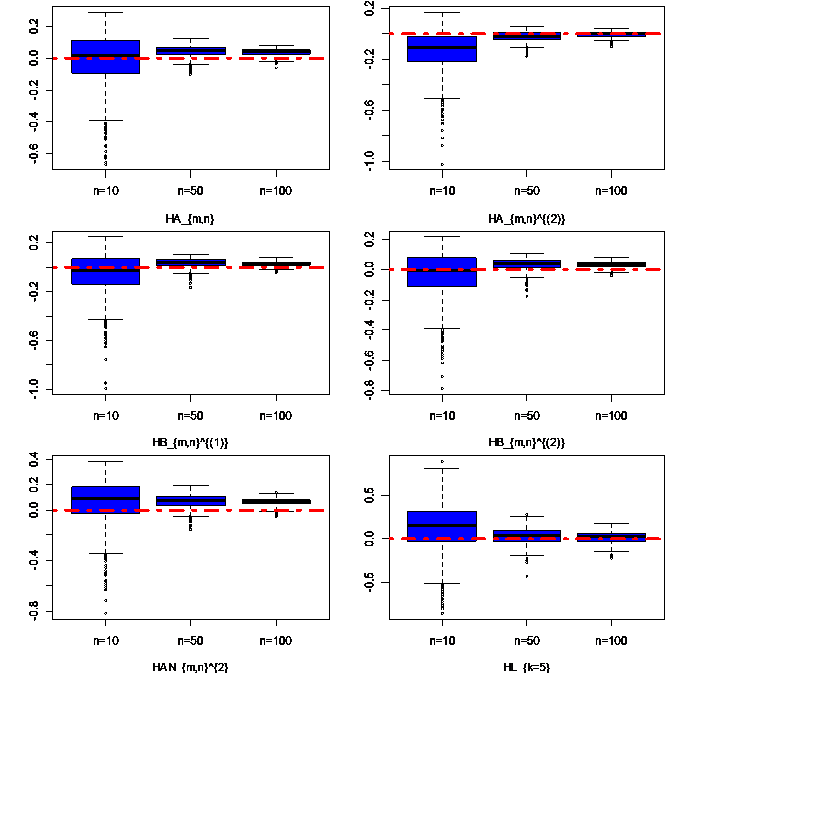}

\caption{Boxplots of the entropy estimates of the four general least RMSE and
$\left|bias\right|$ estimators under the uniform distribution, $n=10,\thinspace50,\thinspace100$}

\end{figure}

\begin{figure}
\includegraphics{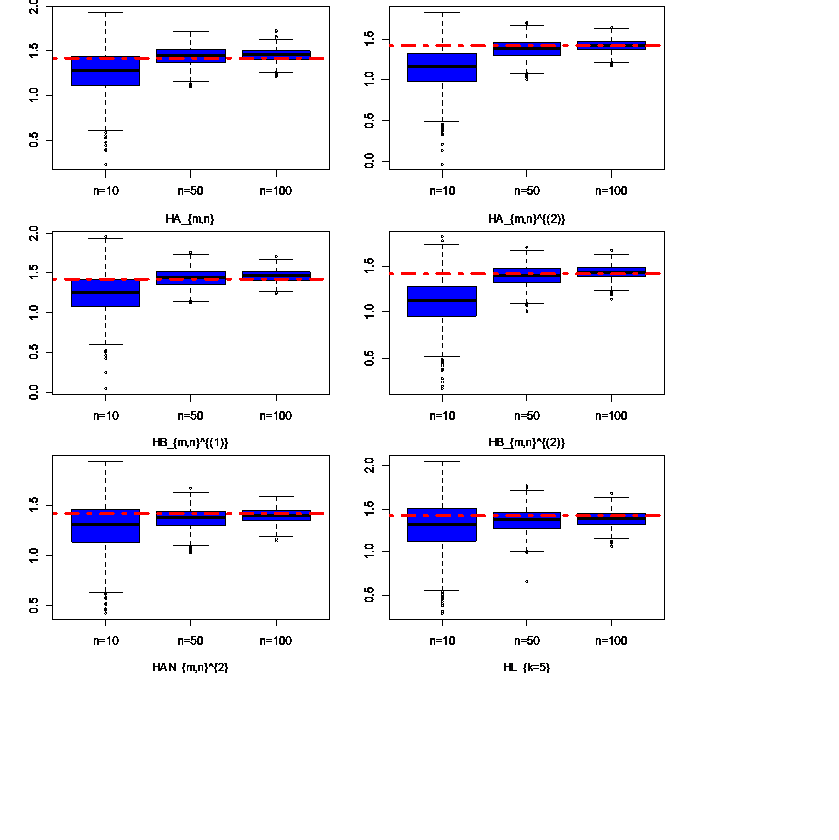}

\caption{Boxplots of the entropy estimates of the four general least RMSE and
$\left|bias\right|$ estimators under the normal distribution, $n=10,\thinspace50,\thinspace100$}

\end{figure}

\begin{figure}
\includegraphics{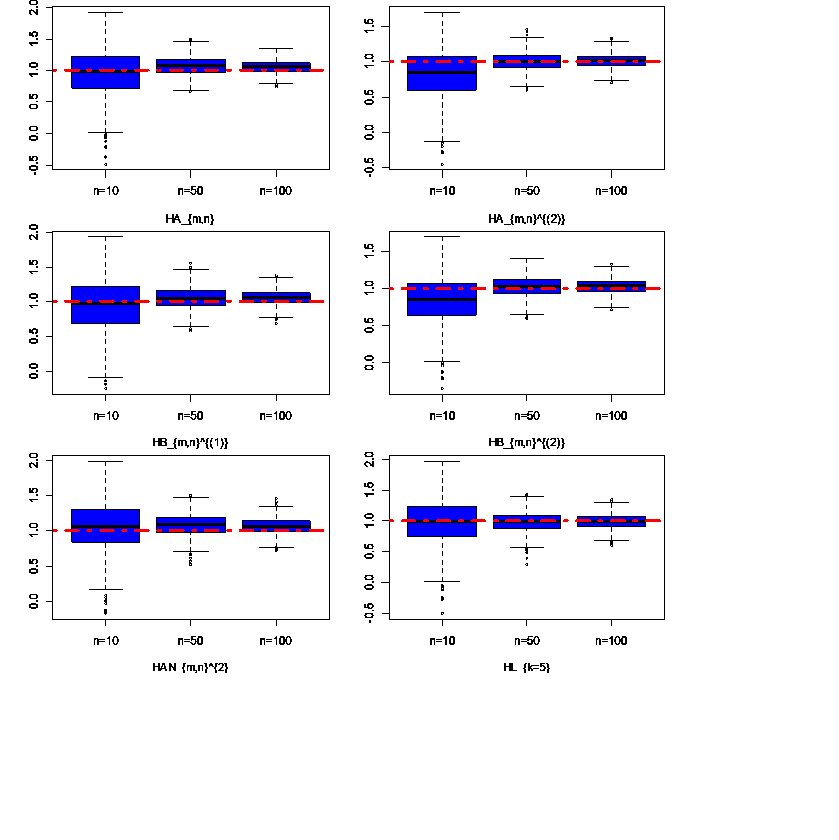}

\caption{Boxplots of the entropy estimates of the four general least RMSE and
$\left|bias\right|$ estimators under the exponential distribution,
$n=10,\thinspace50,\thinspace100$}

\end{figure}

For the results at $d=2,\thinspace3,\thinspace5$, as contained in
Table 12, it could be seen that the $HV$ recorded the least performance
among the four best estimators under this category of multivariate
entropy estimators.

\section{Conclusion}

This work has reviewed several estimators of the Shannon differential
entropy obtained from three main classes, namely: the window size
spacings, kernel density estimation and $k-$nearest neighbour. Also,
27 of them across the three classes have been compared through extensive
simulation studies by comparing their RMSE and $\left|bias\right|$
values at sample sizes $n=10,\thinspace50$ and 100 under the uniform,
normal and exponential distributions at both univariate and multivariate
levels ($d=1,\thinspace2,\thinspace3,\thinspace5$), where multivariate
comparison was possible only with estimators that were not functions
of the window size, $w$.

For the spacings based estimators in both the spacings and KDE classes,
it was discovered that there was no optimal window size for all the
estimators. As a result, an optimal set of window sizes, $m^{+}=m^{*}\pm2$,
where $m^{*}\thinspace=\thinspace\left[\sqrt{n}\thinspace+\thinspace0.5\right]$,
was obtained. Also, no paricular estimator was generally the best
in terms of RMSE and bias values in this class across all the sample
sizes and distributions. However, it is suggested in this work that
the $HA_{m,n}$, $HA_{m,n}^{(2)}$, $HB_{m,n}^{(1)}$, and $HB_{m,n}^{(2)}$
performed very well and should be recommended as good spacings based
estimators of the Shannon entropy irrespective of sample size and
distribution under consideration. Again, the $HAL_{d=1},$$HAN_{m,n}$
and $HAN_{m,n}^{2}$ were observed to have best RMSE and bias values
among all the KDE based estimators in all the sample sizes and distributions
under univariate consideration. However, they were inferior to the
spacings based generally best estimators, except for the $HAN_{m,n}^{2}$
which was almost at par with them. As a result, the $HAN_{m,n}^{2}$
may also be recommended as a good estimator of the Shannon entropy
irrespective of the sample size and distribution. Furthermore, the
$k$NN based estimators were generally weak estimators of the Shannon
differential entropy when compared with those of the other two classes
irrespective of the sample size and distribution under consideration.
The only advantage of the estimators in this class over those of the
other classes was their tractability in any dimension, $d=1,\thinspace2,\thinspace.\thinspace.\thinspace.$.
Also, there was no best number of neighbours, $k$, for all the estimators
in this class at different sample sizes, dimensions and distributions.
However, $k=1,\thinspace2,\thinspace3,\thinspace4,\thinspace5$ may
be used but recommended to approach $k=1$ as $d$ increases. The
$HKL$ is not recommended for use in $d\geq2$. This is because it
lacked consisteny property of an estimator at high dimensions, as
its RMSE and bias values increased with increasing sample size at
$d\geq2$.

Finally, the biases of the estimators vanished asymptotically, with
decreasing asymptotic variances. Also, some estimators maintained
asymptotic symmetric distributions while others maintained asymptotic
skewed distributions with various degrees and types (left/right skewed)
of skewness across the parent distributions.

\section*{Acknowledgements}
\noindent The authors wish to thank TWAS and UNESCO, for granting the first-named author TWAS-UNESCO Associateship Scheme. His research stay at the Institute of Mathematics, VAST, Hanoi was funded by UNESCO.

\end{document}